\documentclass[10pt, journal, compsoc, final]{IEEEtran}

\usepackage{subfigure}

\usepackage{url}

\usepackage{amstext,amsmath,amssymb,amsfonts}
\usepackage{algorithmic,algorithm}
\usepackage[autostyle]{csquotes}  
\usepackage{mathtools,cuted}
\usepackage{color}

\usepackage{booktabs}
\usepackage{tabularx}

\makeatother

\usepackage[backend=bibtex,style=ieee,doi=false, isbn=false, url=false, eprint=false, maxnames=3]{biblatex}
\newcommand{\bs}[1]{\boldsymbol{#1}}

\newcommand{\Vector}[1]{\bs{#1}}
\DeclareMathAlphabet\mathbfcal{OMS}{cmsy}{b}{n}
\newcommand{\Matrix}[1]{\mathbf{#1}}
\newcommand{\Tensor}[1]{\mathbf{\mathcal{#1}}}

\newcommand{\Graph}[1]{\mathit{\mathbf{\bs{#1}}}}
\newcommand{\Set}[1]{\mathit{#1}}

\newcommand{\Hypergraph}[1]{\mathbfcal{#1}}
\newcommand{\Hyperset}[1]{\mathcal{#1}}

\newcommand{\R}{\mathbb{R}}

\newcommand{\where}{\mid}

\begin{document}
\title{Triangular Alignment (TAME): \\ A Tensor-based Approach for \\Higher-order Network Alignment}

\author{Shahin~Mohammadi,~David~F.~Gleich,~Tamara~G.~Kolda, and~Ananth~Grama %
\IEEEcompsocitemizethanks{\IEEEcompsocthanksitem S.~Mohammadi, D.~F.~Gleich, and A.~Grama are with the Department of Computer Science, Purdue University, West Lafayette, IN 47907\protect\\
E-mail: \{mohammadi, dgleich, ayg\}@purdue.edu.
\IEEEcompsocthanksitem T.~G.~Kolda is with the Sandia National Laboratories, Livermore, CA 94551.\protect\\
E-mail: tgkolda@sandia.gov.}
}



\IEEEtitleabstractindextext{\begin{abstract} 

Network alignment has extensive applications in comparative interactomics. Traditional approaches aim to simultaneously maximize the number of conserved edges and the underlying similarity of aligned entities. We propose a novel formulation of the network alignment problem that extends topological similarity to higher-order structures and provides a new objective function that maximizes the number of aligned substructures. This objective function corresponds to an integer programming problem, which is NP-hard. Consequently, we identify a closely related surrogate function whose maximization results in a tensor eigenvector problem. Based on this formulation, we present an algorithm called Triangular AlignMEnt (TAME), which attempts to maximize the number of aligned triangles across networks. 
Using a case study on the NAPAbench dataset, we show that triangular alignment is capable of producing mappings with high node correctness.  We further evaluate our method by aligning yeast and human interactomes. Our results indicate that TAME outperforms the state-of-art alignment methods in terms of conserved triangles. In addition, we show that the number of conserved triangles is more significantly correlated, compared to the conserved edge, with node correctness and co-expression of edges. Our formulation and resulting algorithms can be easily extended to arbitrary motifs.
\end{abstract}

}

\maketitle

\IEEEdisplaynontitleabstractindextext

%
\IEEEpeerreviewmaketitle

\section{Introduction}\label{sec:introduction}



\IEEEPARstart{M}{odeling} cellular machinery as a network of interacting biomolecules
provides significant opportunities for understanding and controlling various biological
processes. This complex network, or \textit{interactome}, may include direct relationships
among biomolecules, such as physical, regulatory, or signaling interactions, or
indirect phenotypic relationships such as epistatic interactions. One common 
abstraction is a protein-protein interaction network (PPI), which is an 
undirected graph, where nodes represent proteins and
edges encode physical interactions among pairs of proteins.
Protein-protein interaction networks are extensively used for modeling and understanding
pathways and protein complexes with respect to their organization and function.

Network motifs are one means of identifying organization and function in these networks.
A network motif is a connected subgraph that occurs
with significantly higher frequency compared to an ensemble of random
graphs with the same size and degree distribution. Over-representation of these patterns are hypothesized to be related to their functional significance~\cite{Milo};
and, indeed, network motif analysis uncovers fundamental circuits that are repeatedly used
to perform critical functions within the cell. 
Due to their important role in decoding
biological networks, various algorithms have been proposed in literature for network
motif detection~\cite{Mfinder, Pajek, MAVisto,  FANMOD, Kavosh}. 
These methods have
identified key motifs, such as feed forward and feedback cycles in directed networks
and triangles in undirected graphs. Furthermore, these motifs are shown to be
involved in regulating cell function, as well as influencing global network
characteristics~\cite{Mangan2003,Shen-Orr2002,Chung2015,Barabasi2004,Wuchty2003}.


Concurrent with the development of methods for network motif detection,
there has been ongoing work on \textit{network alignment} algorithms
for identification of conserved modules across networks. The goal of network alignment
is to identify a mapping between nodes of networks that maximizes similarity (as
defined by a suitable measure) between mapped entities. These mappings can be used to infer
orthologies for unannotated proteins, as well as transferring known biology regarding common
pathways, protein complexes, recurring building blocks, and missing interactions.
These conserved substructures are important since cellular functions require
all of their constituent components (nodes and their interactions) to be conserved.
Conversely, conserved modules provide insights into corresponding functional
organization~\cite{molecular2modular}.

The network alignment problem, unlike its counterpart over sequences, is NP-complete to
solve exactly, since in the most generic form it can be reduced to the subgraph
isomorphism problem. However, heuristics have been proposed 
to generate useful answers through various reformulations as well as
by incorporating additional data to guide the alignment process. We survey
these methods in Section~\ref{sec:background} in more detail. 
An important distinction among alignment methods is their local versus global nature. Local alignment methods aim to identify conserved functional modules between networks, such as signaling pathways and protein complexes, by optimally aligning these substructures. Due to duplication-divergence events, there can be more than one match for each substructure, which makes alignments ambiguous. Global network alignment, on the other hand, aims to identify a one-to-one mapping between all pair of nodes in the input graphs that maximizes similarity of aligned nodes.

In this paper, we introduce a new class of methods based on \textit{higher-order network
alignment}. These methods combine strengths of both global and local network alignment.
In this framework, users can define any network motif structure of interest to drive the
alignment process. These general structures can be represented through a \textit{motif tensor},
in which the order of tensor is the same as the size of the given subgraph template.
We encode the higher order network alignment using a tensor-based formulation and show that
the exact solution to the alignment problem is equivalent to solving a higher-order
integer program, which is NP-hard. To optimize this objective function on large networks,
we exploit a bijection between
the eigenpairs of the motif tensor and a heuristic approximation of the 
integer program. We can then use the previously proposed SS-HOPM~\cite{SSHOPM} method
to identify maximizing dominant eigenpairs of a symmetric tensor,
and propose a higher-order alignment method based on this scheme. The motif tensor of the
alignment graph can be represented as the Kronecker product of motif tensors for each input graph.
This tensor is too large to fit in the memory for typical PPI networks, even for small motifs.
We develop a novel implicit kernel for computing the tensor-vector product as the main
operator within SS-HOPM.  Similar kernels have been previously proposed in the context of computer vision research
~\cite{svab2007-exploiting,Chertok-2010-higher-order}; however, we present a highly efficient, motif-centered version that is easily extensible to higher order sub-structures.

Using a case study of triangle motifs, we present a complete algorithm, called
\textit{Triangular AlignMEnt (TAME)}. 
We propose a constrained variant of our algorithm, \textit{cTAME}, that
operates only on a subset of reliable nodes. This method provides better accuracy in cases where 
sequence similarity scores are highly reliable proxies for the true-positive alignments. We further validate our method and show that it compares favorably to the state-of-art methods for network alignment on both synthetic datasets from NAPAbench~\cite{NAPAbench} and alignment of yeast and human interactomes. Our framework can be easily extended to
arbitrary subgraphs and motivates an alternate view to the network alignment problem.

\section{Background and previous methods}
\label{sec:background}

Based on the alignment strategy, we can generally classify different methods as
either \textit{local} or \textit{global} alignment techniques. Local alignment aims
to identify common substructures corresponding to pathways or protein complexes
that are conserved in networks  of different species.
Local alignments often yield ambiguous
mappings, since functional building blocks can have many-to-many relationships. On
the other hand, global alignment attempts to find the best overall mapping between
the nodes of input graphs that maximizes both functional and topological similarity
of aligned nodes, while enforcing the one-to-one constraint. In pairwise alignment,
this leads to a unique alignment for each node in the smaller graph to a 
node in the larger graph. These alignments are unambiguous and can be used
to transfer functional orthologies between pairs of proteins, as well as to compute
the overall similarity of input graphs.

Local aligners differ in the topology of the substructures they search for, their objective formulation, and search strategy. PathBlast~\cite{PathBlast1, PathBlast2} and NetworkBlast~\cite{NetworkBlast, NetworkBlast-M} 
are early examples that use a probabilistic scoring function to search for linear paths and clique-like substructures, respectively. Flannick \textit{et al.}~\cite{Graemlin1} proposed an evolutionarily-motivated scoring function and incorporated known alignments via a supervised learning scheme~\cite{Graemlin2}.
Koyuturk \textit{et al.}~\cite{MaWish1, MaWish2} instead posed local alignment 
as a suitably formulated optimization problem in their MaWISh framework. AlignMCL~\cite{AlignMCL} combines input networks to construct an alignment graph and uses the Markov CLustering (MCL) algorithm to partition the alignment graph and identify protein clusters.

Global aligners, on the other hand, aim to find \textit{unique} node-to-node mappings that maximize a given objective function.
IsoRank ~\cite{IsoRank1, IsoRank2} and its predecessor IsoRankN~\cite{IsoRankN} are among the early methods in this class. The main idea behind IsoRank is that a pair of nodes corresponds to a good match if the nodes are homologous and their respective neighborhoods are similar. This recursive scheme is then cast as an eigenvalue problem, the solution of which is identified using power method. Later, Kollias \textit{et al.} \cite{NSD,FastAlign} proposed methods to speedup similarity computation and matching phases in IsoRank, respectively. GRAphlet-based ALigner (GRAAL)~\cite{GRAAL} was proposed as the first member in a family of ``GRAAL-based methods.'' These methods are distinguished by their use of \textit{graphlet degrees} of nodes, or the number of induced subgraphs with given topology incident on each vertex, as a topological signature for vertices. \textit{H-GRAAL} is an extension of the GRAAL method that uses the Hungarian algorithm, hence the name \textit{H-GRAAL}~\cite{H-GRAAL}, to compute maximum-weight bipartite matching and extract alignments from the node similarity scores. Matching-based Integrative GRAAL (MI-GRAAL) \cite{MI-GRAAL} allows simultaneous \textit{integration} of five different similarity matrices, including similarity of sequences as well as graphlet degrees. Moreover, it uses a novel seed-and-extend \textit{matching} strategy that is shown to outperform the Hungarian method in terms of alignment quality. C-GRAAL~\cite{C-GRAAL} is a ``common-neighbors-based'' addition to the GRAAL family that introduces additional heuristics to the seed-and-extend matching strategy. It uses the concept of \textit{neighborhood densities} to choose the best matching pairs to align. Both heuristics proposed in \textit{MI-GRAAL} and \textit{C-GRAAL} aim to implicitly maximize the number of aligned/conserved edges using a greedy alignment strategy. Klau \textit{et al.}~\cite{Natalie1,Natalie2} formulate the network alignment problem as an integer quadratic problem (IQP) that aims to simultaneously maximize a convex combination of the total  number of aligned edges and overall similarity of mapped nodes. They proposed a method, named NATALIE, which uses Lagrangian relaxation to provide a real approximation to the network alignment problem formulated as an IQP. Bayati \textit{et al.}~\cite{Bayati-2013-netalign} proposed a message passing algorithm to solve the IQP proposed by Klau \textit{et al.} by utilizing the sparsity pattern in the sequence similarity search space. More recently, the latest addition to the GRAAL-family, called Lagrangian GRAAL (L-GRAAL)~\cite{L-GRAAL}, has been proposed. L-GRAAL is similar to previous members in that it uses graphlet degree signatures as a source of topological similarities. However, it uses a seed-and-extend step that is based on the integer programming and Lagrangian relaxation similar to the one proposed in NATALIE.

In addition to the GRAAL-family, many other methods have been developed. GHOST~\cite{GHOST} uses a spectral approach that, similar to GRAAL, tries to encode local topology of each node using a signature vector, called the \textit{spectral signature}. The similarity of nodes is then identified based on the similarity of their spectral signature. GHOST uses local swaps, similar to the ones proposed in PISwap~\cite{PISWAP}, to post-process the final alignment and improve the results. Similarly, MAGNA \cite{MAGNA} is an evolutionary algorithm designed to enhance alignments computed by other methods. The method can also be used as an independent aligner by applying it to a random initial population. MAGNA++ \cite{MAGNA++} is an extension of MAGNA that allows integration of sequence similarities and provides a GUI for MAGNA. More recently, Gong \textit{et al.}~\cite{Gong2015} proposed a memetic algorithm-based algorithm for alignment which is similar in nature to MAGNA in that it iteratively updates alignment using evolutionary swapping operators.
HubAlign \cite{HubAlign} is a recent network aligner that is based on similar concepts as MaWISh \cite{MAWISH}. It uses a minimum degree heuristic to identify and align ``important" proteins first, and to use them as anchors to locally extend the alignment to the whole network. Mohammadi \textit{et al.}~\cite{AlignmentChapter}, Clark \textit{et al.}~\cite{Clark2014}, and Elmsallati \textit{et al.}~\cite{Elmsallati2015} provide a comprehensive comparison of these methods.

\section{Notations and Terminology}
\subsection{Graphs and Hypergraphs}

Biochemical networks are often modeled as graphs in which vertices (or nodes) represent
biomolecules (proteins, genes, etc.) and edges (or arcs) encode pairwise relationships among
them. Formally, a graph $\Graph{G}$ is represented by $\Graph{G} = (\Set{V_G}, \Set{E_G})$,
where $\Set{V_G}$ is a finite set of vertices, $\Set{V_G} = \{v_1, v_2, \ldots., v_n\}$, and
$\Set{E_G}$ is a finite set of edges, denoted by $(v_i, v_j)$, such that $\Set{E_G} \subseteq
(\Set{V_G} \times \Set{V_G})$. We focus on undirected graphs, where edges define a 
symmetric relation among graph vertices.
A graph can be represented by a matrix $\Matrix{A}_\Graph{G}$ of size $|\Set{V_G}|\times|\Set{V_G}|$,
known as the \textit{adjacency matrix}, in which $\Matrix{A}_G(i, j) = 1$ when $(v_i, v_j) \in
\Set{E_G}$. The graph neighborhood for each node $v_i$ in the graph, represented by $\Set{N}_G(i)$,
is defined as the set of nodes that have an edge with $v_i$; formally $\Set{N}_G(i) = \{j \where (i, j)
\in \Set{E_G} \}$. Given a pair of graphs, $\Graph{G} = (\Set{V_G}, \Set{E_G})$ and
$\Graph{H} = (\Set{V_H}, \Set{E_H})$, their Kronecker product is a graph with $|V_G| \times |V_H|$
vertices that are formed by pairs of vertices from $\Graph{G}$ and $\Graph{H}$, e.g., $ii'$ for $i$ from $\Set{V_G}$ and $i'$ from $\Set{V_H}$. and where each edges between nodes $ii'$ and $jj'$ corresponds to $(i, j) \in E_G$ and
$(i', j') \in E_H$.

Hypergraphs are natural generalizations of graphs in which the relations among vertices is
not restricted to be pairwise. Formally, a hypergraph is defined using the pair
$\Hypergraph{G}=(\Set{V}_G, \Hyperset{E}_G)$, where $\Set{V}$ is the set of vertices and
$\Hyperset{E}$ is the set of \textit{hyperedges}. Here, each hyperedge defines a relationship
among a nonempty subset of vertices. A \textit{$k$-uniform} hypergraph is a hypergraph in
which the cardinality of each hyperedge is exactly $k$. As such, 2-uniform hypergraphs are
equivalent to traditional graphs. A $k$-uniform hypergraph can be represented by a $k^{th}$-order
tensor $\Tensor{T}_G$, known as the \textit{adjacency tensor}, where
$\Tensor{T}_G(i_1, i_2, \ldots, i_k) = 1$ iff $(i_1, i_2, \ldots, i_k) \in \Hyperset{E}_G$.
The hypergraph incidence set for each node $v_i$ in a $k$-uniform hypergraph, represented by
$\Hyperset{N}_G(i)$, is the set of $(k-1)$-node subsets such that adding node $i$ to each
subset forms an edge. This is one possible generalization of a neighborhood to a hypergraph,
and is formally defined as $\Hyperset{N}_G(i) = \{(i_2, i_3, \ldots, i_k) \where (i, i_2, i_3, \ldots, i_k) \in
\Hyperset{E}_G \}$. For a given graph $\Graph{G} = (\Set{V_G}, \Set{E_G})$ and a given size
$k$ structural motif $\Graph{M} = (\Set{V_M}, \Set{E_M})$, where $|\Set{V_M}| = k$, we can
represent the occurrences of $\Graph{M}$ in $\Graph{G}$ by a $k$-way tensor $\Tensor{T}_G$,
referred to as the \textit{motif-tensor}, where $\Tensor{T}_G(i_1, \ldots, i_k) = 1$ when the
induced subgraph among vertices $\{v_{i_1}, \ldots, v_{i_k}\}$ in $\Graph{G}$ is isomorphic to
$\Graph{M}$. Throughout this paper, we make extensive use of a special case of the motif-tensor
where the substructure of interest is the triangle motif. Given an undirected graph
$\Graph{G}$, its \textit{triangle tensor}, denoted by $\Tensor{\bigtriangleup}_G$, is a
third-order tensor such that $\Tensor{\bigtriangleup}_G(i, j, k) = 1$ iff  $(v_i, v_j),
(v_j, v_k), (v_k, v_i) \in \Set{E}_G$.

\subsection{Tensor definition and properties}
A real-valued $m^{th}$-order $n$-dimensional tensor, denoted by $\Tensor{T}^{[m, n]}$, is a
multiway array, whose entries can be indexed using an $m$-dimensional tuple, and each tensor
way (or mode) has dimension $n$. An $n$-dimensional vector and square matrix are examples
of $1^{st}$-order and $2^{nd}$-order tensors, respectively. We refer to elements of a
tensor $\Tensor{T}^{[m, n]}$ using $\Tensor{T}(i_1, i_2,\ldots, i_m)$ and
$\Tensor{T}_{i_1, i_2,\ldots, i_m}$, interchangeably. A tensor $\Tensor{T}$ is \textit{symmetric} iff:
\begin{equation}
\Tensor{T}(i_1, i_2,\ldots, i_m) = \Tensor{T}(i_{\pi(1)}, i_{\pi(2)},\ldots, i_{\pi(m)})
\end{equation}
for all $\pi \in \Pi_m$, where $\Pi_m$ is the set of all permutations of $(1,2, \ldots, m)$. 
As an example, note that the triangle tensor is a symmetric tensor. 

Given a symmetric tensor $\Tensor{T}^{[m,n]}$, together with an $n$-dimensional vector
$\Vector{x}\in \R^{n}$, we concern ourselves with two operations. The first is the 
tensor-vector product: $\Tensor{T} \Vector{x}^{m-1}$, which is defined element-wise as
\begin{equation} \label{eq:ttv}
  (\Tensor{T} \Vector{x}^{m-1})_{i_1} = \sum_{i_2=1}^n \sum_{i_3=1}^{n} \cdots \sum_{i_m=1}^{n} \Tensor{T}_{i_1, i_2,\ldots, i_m} x_{i_2}   x_{i_3} \ldots  x_{i_m}.
\end{equation}
The second is the scalar polynomial form in $\Vector{x}$: $\Tensor{T} \Vector{x}^{m}$ defined as
\begin{equation}
 \Tensor{T} \Vector{x}^{m} = \sum_{i_1=1}^n \sum_{i_2=1}^n \sum_{i_3=1}^{n} \cdots \sum_{i_m=1}^{n} \Tensor{T}_{i_1, i_2,\ldots, i_m} x_{i_1} x_{i_2}   x_{i_3} \ldots x_{i_m}.
 \label{eq:contract}
\end{equation}
Note that $\Vector{x}^T (\Tensor{T} \Vector{x}^{m-1}) = \Tensor{T} \Vector{x}^{m}$.


A pair $(\lambda, \Vector{x})$, $\lambda \in \R, \Vector{x} \in \R^n$, is the Z-eigenpair of
the symmetric tensor $\Tensor{T}$ when:
\begin{equation}
\Tensor{T}\Vector{x}^{m-1} = \lambda \Vector{x}; \text{ with } \| \Vector{x} \|_2 = 1.
\label{eq:eigen}
\end{equation}
Any eigenpair $(\lambda, \Vector{x})$ of tensor $\Tensor{T}$ is a Karush-Kuhn-Tucker (KKT)
point of the following nonlinear optimization problem~\cite{Lim}:
\begin{equation}
\begin{array}{ll}
 \displaystyle \mathop{\text{maximize}}_{\Vector{x} \in \R^n} & \Tensor{T} \Vector{x}^m \\
 \text{subject to} & \| \Vector{x} \|_2 = 1. 
\end{array}
\label{eq:optim}
\end{equation}

We will use one additional concept: the Kronecker product of tensors. This type of product is a reshaped
version of the well-established outer product for tensors. 
Formally, given a pair of symmetric tensors, $\Tensor{T}_1 \in \mathcal{R}^{[m, n_1]}, \Tensor{T}_2 \in
\mathcal{R}^{[m, n_2]}$, their Kronecker product, denoted by $\Tensor{T}_1 \otimes \Tensor{T}_2 \in
\mathcal{R}^{[m, n_1n_2]}$, is defined as:
\begin{equation}
(\Tensor{T}_1 \otimes \Tensor{T}_2)(i_1^{}i_1',  \ldots, i_m^{}i_m') = \Tensor{T}_1 (i_1, \ldots, i_m) \Tensor{T}_2 (i_1', \ldots, i_m')
\end{equation}
with $1 \leq i_1, \ldots, i_m \leq n_1$, $1 \leq i_1', \ldots, i_m' \leq n_2$, and the notation $i_ki'_k$ denotes a specific index for the index representing the pair which is $(i_k - 1)n_2 +i'_k$. 

%
%

\section{Higher-Order Network Alignment}

Our framework for higher-order network alignment draws heavily on the integer quadratic program
for global network alignment. We begin by reviewing this formulation~\cite{BP,Natalie1,Natalie2}.

\subsection{Formulation of global network alignment as a Binary Quadratic Program (BQP)}
\label{sec:GNAnotations}

Given a pair of networks, represented by $\Graph{G}=(\Set{V}_G, \Set{E}_G)$ and
$\Graph{H}=(\Set{V}_H, \Set{E}_H)$,
the global network alignment problem aims to find an optimal one-to-one mapping between vertices of
$\Graph{G}$ and $\Graph{H}$ that maximizes both the prior (known) similarity and the topological similarity among pairs of aligned nodes.  The topological similarity is the number of \emph{edges} preserved in both $\Graph{G}$ and $\Graph{H}$ under the matching.

Let us denote by $\Vector{w}(ii')$ the prior similarity of a pair of nodes $i \in \Set{V}_\Graph{G}$ and $i' \in \Set{V}_\Graph{H}$. Here, $ii'$ is a shorthand for the linear index $(i' - 1)|\Set{V}_\Graph{G}| + i$. (For more detail about this derivation, please see~\cite{BP}.) Furthermore, we use a binary indicator vector $\Vector{x}$ to represent matches, where $\Vector{x}(ii')$ is one if node $i$ is matched to node $i'$, and zero otherwise. Finally, we define a binary matrix $\Matrix{S}$, where $\Matrix{S}(ii', jj')$ is one if $(i, j)  \in \Set{E}_\Graph{G}$ \textbf{and} $(i', j') \in \Set{E}_{\Graph{H}}$, and zero otherwise. For the choice of $ii'$ listed above, $\Matrix{S} = A_{\Graph{H}} \otimes A_{\Graph{G}}$. Using this notation, we can write the global network alignment as the following binary quadratic program (BQP):
\begin{equation}
 \begin{array}{ll}
  \displaystyle \mathop{\text{maximize}}_{\Vector{x}} & \displaystyle (1-\alpha) \Vector{w}^T \Vector{x} + \frac{\alpha}{2} \Vector{x}^T \Matrix{S} \Vector{x} \\
 \text{subject to} & \Matrix{C} \Vector{x} \le \Vector{1}_{|V_G|+|V_H|} \\
  & \Vector{x}(ii') \in \{0, 1\}.
 \end{array}
 \label{eq:IQP}
\end{equation}
where $\Matrix{C}$ is the unsigned node-edge incidence matrix of the complete bipartite
graph on $|V_G|$ and $|V_H|$ vertices. This problem is also equivalent to a binary linear program through a standard linearizing transformation~\cite{Natalie1,BP}. Different global alignment methods can be viewed as algorithms that either implicitly or explicitly 
optimize this BQP formulation.

\subsection{The higher-order generalization}

We generalize the node alignment solutions proposed earlier to the problem of aligning
higher-order substructures in graphs. As previously mentioned, 
this is motivated by the existence of motifs in
biological networks. Again, we will use vectors whose indices represent pairs $ii'$ as above. 
Let $\Tensor{T}_G$ and $\Tensor{T}_H$ be the motif-tensors associated with a motif $\Graph{M}$
in both graphs $G$ and $H$, where this motif has $m$-nodes. Then the 
higher-order network alignment problem is the binary polynomial problem:
\begin{equation}
 \begin{array}{ll}
  \displaystyle \mathop{\text{maximize}}_{\Vector{x}} & \displaystyle (1-\alpha) \Vector{w}^T \Vector{x} + \frac{\alpha}{m!} (\Tensor{T}_H \otimes \Tensor{T}_G) \Vector{x}^{m} \\
 \text{subject to} & \Matrix{C} \Vector{x} \le \Vector{1}_{|V_G|+|V_H|} \\
  & \Vector{x}(ii') \in \{0, 1\}.
 \end{array}
 \label{eq:HONA-BP}
\end{equation}
This problem can again be converted into a binary linear program through a standard
linearization procedure on the higher-order polynomials, but that is tangential
to our discussion here. As a generalization of the network alignment problem,
this problem is NP-hard as well.
 
We focus on \textit{triangle} motifs,
which are special cases of feed-forward/backward motifs. Please note that our method
is general, and can be applied to arbitrary motifs.
Denote the triangle tensor of graph $\Graph{G}$ and $\Graph{H}$ using 
$\Tensor{\bigtriangleup}_G$ and $\Tensor{\bigtriangleup}_H$, respectively. Also, denote the 
triangle tensor for the product graph by $\Tensor{\bigtriangleup}_{H \times G} = 
\Tensor{\bigtriangleup}_H \otimes \Tensor{\bigtriangleup}_G$. Using this notation, we can write 
the higher-order network alignment problem as a binary cubic program:
\begin{equation}
 \begin{array}{ll}
  \displaystyle \mathop{\text{maximize}}_{\Vector{x}} & \displaystyle (1-\alpha) \Vector{w}^T \Vector{x} + \frac{\alpha}{6} (\Tensor{\bigtriangleup}_{H \times G}) \Vector{x}^{3} \\
 \text{subject to} & \Matrix{C} \Vector{x} \le \Vector{1}_{|V_G|+|V_H|} \\
  & \Vector{x}(ii') \in \{0, 1\}.
 \end{array}
 \label{eq:TAME}
\end{equation}
In this formulation, $\Tensor{\bigtriangleup}_{H \times G} \Vector{x}^3$ counts the number of 
triangles that are conserved under the alignment represented by vector $\Vector{x}$, and 
$\Vector{w}^T\Vector{x}$ plays a similar role as in BQP formulation of global alignment. 

We propose a heuristic procedure to optimize this objective. 
First, we remove the one-to-one constraint on $\Vector{x}$ from the
optimization problem. Second, we relax the the problem over the reals. In this case, the solution is unbounded. So we introduce a 2-norm constraint on the solution vector $\Vector{x}$. When $\alpha = 1$, then the resulting problem coincides with the 
eigenvectors of tensor $\Tensor{\bigtriangleup}_{H \times G}$, as presented in 
Equation~\ref{eq:optim}.  Specifically, the surrogate we use is:
\begin{equation}
\label{eq:surrogate}
 \begin{array}{ll}
  \displaystyle \mathop{\text{maximize}}_{\Vector{x}} &  (\Tensor{\bigtriangleup}_{H \times G}) \Vector{x}^{3} \\
 \text{subject to} & \|\Vector{x}\| = 1.
 \end{array}
\end{equation}
For this problem, there is a known algorithm, called \textit{shifted 
symmetric higher-order power method (SS-HOPM)}~\cite{SSHOPM}, which can be used to identify eigenpairs of 
$\Tensor{\bigtriangleup}_{H \times G}$ with large eigenvalues. 
When $\alpha$ is not 1, we still compute the same tensor eigenvector.
We incorporate sequence similarities 
encoded by $\Vector{w}$ by starting iterations from $\Vector{x}_0 = \Vector{w}$.
Finally,
we take the real-valued solution from SS-HOPM and form a matrix 
$\Matrix{X}(i,i')=\Vector{x}(ii')$ that we will post-process to produce a 1-1 matching.

We now describe a number
of ways to exploit the structure of the problem in this setup for a more efficient implementation. The fundamental computational difficulty is manipulating the tensor
$\Tensor{\bigtriangleup}_{H \times G}$. A straightforward
application of SS-HOPM on this tensor would utilize a data structure that consumes
55 TB of memory (this exploits sparsity alone, like traditional graph algorithms). 
To see how this structure balloons in size, consider the case of aligning the yeast and 
human interactomes. These networks have $347,079$ and $407,650$ triangles, respectively.
If we store the non-zeros in a sparse tensor representation of $\Tensor{\bigtriangleup}_{H \times G}$ using three 32-bit
indices per non-zero, it requires (347,079$~\times~$ 407,650 motifs)$~\times~$(36 symmetry non-zeros per motif)$~\times~$(12 bytes per non-zero)$~\approx~$55.5 terabytes of memory to store the product tensor ( or 1.5$ \text{ terabytes}$, if we exploit the symmetry).
Forming the full non-zero structure, however, is unnecessary as we only need to use this tensor
to compute the tensor-times-vector operation. This can be done implicitly without
forming the complete structure, which is discussed in Section~\ref{sec:impTTV}.

\subsection{An implicit kernel for computing tensor-vector products}
\label{sec:impTTV}

The key challenge in computing the tensor-vector product is that the number of elements in the triangle tensor of product graph, $\Tensor{\bigtriangleup}_{H \times G}$, is too large to fit in the memory of most modern computers, even for relatively small graphs. To remedy this problem, we note that there is no need to explicitly construct $\Tensor{\bigtriangleup}_{H \times G}$. All we need to run SS-HOPM is to compute $\Tensor{\bigtriangleup}_{H \times G}\Vector{x}^3$ and $\Tensor{\bigtriangleup}_{H \times G}\Vector{x}^2$. Re-writing the tensor-vector product formulation in Equation~\ref{eq:ttv}, we find the following vertex-centered, implicit kernel as follows:

\begin{equation}
\begin{aligned}
& (\Tensor{\bigtriangleup}_{H \times G}\Vector{x}^2)_{ii'} \\
& \quad = \sum_{\mathclap{jj', kk'}} \Tensor{\bigtriangleup}_{H \times G} (ii', jj', kk') \Vector{x}(jj')\Vector{x}(kk') \\
& \quad = \sum_{\mathclap{j, j', k, k'}} \Tensor{\bigtriangleup}_G(i, j, k)\Tensor{\bigtriangleup}_H(i', j', k') \Matrix{X}(j, j')\Matrix{X}(k, k')\\ 
& \quad = \sum_{\mathclap{j, k}} \Tensor{\bigtriangleup}_G(i, j, k)\sum_{j'} \Matrix{X}(j, j')\sum_{k'}\Tensor{\bigtriangleup}_H(i', j', k') \Matrix{X}(k, k')
\label{eq:impKernel_vectex}
\end{aligned}
\end{equation}
where $\Matrix{X} = \textbf{unvec}(\Vector{x})$. Additionally, we can simplify this 
vertex-centered formulation to derive a more efficient motif-centered kernel. To this end, we note 
that triangle tensors of graphs $\Graph{G}$ and $\Graph{H}$ represent a 3-uniform hypergraph over 
the set of vertices $V_G$ and $V_H$, respectively. Denote the hypergraph incidences of 
these hypergraphs by $\Hyperset{N}_{\bigtriangleup_G}$ and $\Hyperset{N}_{\bigtriangleup_H}$, 
where $\Hyperset{N}_{\bigtriangleup_G}(i) = \{(j, k) \where (v_i, v_j), (v_j, v_k), (v_k, v_i) \in 
\Set{E}_G\}$, and $\Hyperset{N}_{\bigtriangleup_H}(i') = \{(j', k') \where (v_{i'}, v_{j'}), (v_{j'}, v_{k'}), 
(v_{k'}, v_{i'}) \in \Set{E}_H\}$. Then
\begin{equation}
 \label{eq:impKernel_motif}
\begin{aligned}
 & \Tensor{\bigtriangleup}_{H \times G}\Vector{x}^2(ii') = \\ & \qquad 2 \sum_{\mathclap{(j, k) \in \Hyperset{N}_{\bigtriangleup_G}(i)}} 
  \textstyle \sum_{(j', k') \in \Hyperset{N}_{\bigtriangleup_H}(i')} \scriptstyle \Matrix{X}(j, j')\Matrix{X}(k, k') + \Matrix{X}(j, k')\Matrix{X}(k, j').\\
\end{aligned}
\end{equation}
In this formulation, we make use of the symmetric property of triangle tensors, where the outer 
factor of $2$ corresponds to the $(m-1)$ degrees of symmetries. The inner summation accounts for the fact that nodes $v_j$ and $v_k$ from $\Graph{G}$, or vertices $\{ v_{i_2},\ldots, v_{i_k} \}$ when dealing with motif-tensors of 
size $k$, can be mapped to their counterpart vertices $v_{j'}$ and $v_{k'}$ in $\Graph{H}$ in 
$(m-1)!$ different ways. Each of these mappings contribute a factor of one in 
$\Tensor{\bigtriangleup}_{H \times G}$. However, their corresponding $\Vector{x}$ values are 
different and we need to separately compute their product. We use this motif-centered formulation 
in our final algorithm to compute the implicit tensor-kernel product, $\tilde{\Vector{x}} = 
\Tensor{\bigtriangleup}_{H \times G}\Vector{x}^{2}$. Having $\tilde{\Vector{x}}$, one can easily 
compute $\Tensor{\bigtriangleup}_{H \times G}\Vector{x}^{3} = \Vector{x}^{T}\tilde{\Vector{x}}$. 
The simplified pseudo-code of the implicit kernel for computing $\Tensor{\bigtriangleup}_{G \times 
H}\Vector{x}^{2}$ is provided in Algorithm~\ref{alg:impTTV}. The computation time of this algorithm is of $O(|{\bigtriangleup}_G|\times|{\bigtriangleup}_H|)$.

\begin{algorithm}
\centering
\begin{algorithmic}[1]
\REQUIRE  Triangle-tensors ${\bigtriangleup}_G, \Tensor{\bigtriangleup}_H$,  for $\Graph{G}$ and $\Graph{H}$; a vector $\Vector{x}$
\ENSURE $\Vector{y} = \Tensor{\bigtriangleup}_{H \times G}\Vector{x}^{2}$
\STATE $\Matrix{X} = \textbf{unvec}(\Vector{x})$
\STATE $\Matrix{Y} = \Matrix{0}$
\FOR{$v_i \in V_G$}
	\FOR{$v_{i'} \in V_H$}
		\FOR{$	\{ (j, k) \in \Set{N}_{\bigtriangleup_G}(i) \}$}
			\FOR{$	\{ (j', k') \in \Set{N}_{\bigtriangleup_H}(i') \}$}		
				\STATE $\Matrix{Y}(i, i') \mbox{\texttt{+=}} \Matrix{X}(j, j')\Matrix{X}(k, k') + \Matrix{X}(j, k')\Matrix{X}(k, j')$		
			\ENDFOR
		\ENDFOR
		\STATE $\Matrix{Y}(i, i') = \Matrix{Y}(i, i')*2$
	\ENDFOR
\ENDFOR
\STATE $\Vector{y} = \textbf{vec}(\Matrix{Y})$
\end{algorithmic}
\caption{Implicit tensor-times-vector product (impTTV)}
\label{alg:impTTV}
\end{algorithm}

\subsection{Triangular AlignMEnt (TAME) algorithm}
\label{sec:TAME_alg}

We now integrate different building blocks introduced earlier to present a 
higher-order alignment method for triangle motifs. This code uses
one additional primitive. The function \textbf{score} solves a
bipartite maximum-weight matching problem (using the Hungarian
method) and returns the total number of 
triangles $t$ aligned by the matching. The pseudocode for TAME 
algorithm is presented in Algorithm~\ref{alg:TAME}. 

\begin{algorithm}
\centering
\begin{algorithmic}[1]
\REQUIRE Triangle tensors $\Tensor{\bigtriangleup}_G, \Tensor{\bigtriangleup}_H$;   Sequence similarities $\Vector{w}$; Shift parameter $\beta$
\ENSURE The best topological scores $\Matrix{X}$ from any iteration
\STATE $k = 0$ \COMMENT{Iteration number}
\STATE $\Vector{w} \leftarrow \Vector{w}/\|\Vector{w} \|$
\STATE $\Vector{x_0} = \Vector{w}$
\STATE $t_0 = 0$
\REPEAT
\STATE $\tilde{\Vector{x}}_{k+1} = \textbf{impTTV}(\Tensor{\bigtriangleup}_G, \Tensor{\bigtriangleup}_H, \Vector{x}_k)$ 
\STATE $\lambda_{k+1} = \Vector{x}_{k}^{T} \tilde{\Vector{x}}_{k+1}^{} $
\STATE $\hat{\Vector{x}}_{k+1} = \tilde{\Vector{x}}_{k+1} + \beta \Vector{x}_{k} $ 
\STATE $\Vector{x}_{k+1} = \frac{\hat{\Vector{x}}_{k+1}}{\| \hat{\Vector{x}}_{k+1} \|}$
\STATE $\Matrix{X}_{k+1} = \bf{unvec}(\Vector{x}_{k+1})$
\STATE $ t_{k+1} =  \bf{score}(\Matrix{X}_{k+1})$
\STATE Update $(\Matrix{X}, t)_{\text{best}}$ to $(\Matrix{X}, t)_{k+1}$ if $t_{k+1} > t_{\text{best}}$
\STATE $k = k+1$
\UNTIL {$\lambda_k - \lambda_{k-1}$ is small or the max iteration is hit}
\RETURN $\Matrix{X}_{\text{best}}$

\end{algorithmic}
\caption{The Triangular AlignMEnt (TAME) algorithm}
\label{alg:TAME}
\end{algorithm}

The overall algorithm takes in the prior similarity and uses that to initialize the SS-HOPM (lines 5-14). $\textbf{impTTV}$ procedure uses the motif-centered, implicit tensor-times-vector kernel proposed in Section~\ref{sec:impTTV}. The SS-HOPM main loop generates a sequence of topological similarity 
matrices. However, the SS-HOPM process generates a sequence to optimize the problem after removing
the two constraints in Equation~\ref{eq:TAME}, namely the integer constraint on $\Vector{x}$ and the one-to-one matching 
constraint over $\Matrix{X}$. 
To enforce these constraints, we perform a matching in each iteration and compute topological score as the total number of aligned triangles. We keep the highest scoring topological matrix as $\Matrix{X}_{\text{best}}$.

In addition to Algorithm~\ref{alg:TAME}, which we refer to as \textbf{full TAME}, we present a 
variant of this algorithm, called \textbf{constrained TAME}, which only matches nodes that 
have \emph{at least one match} suggested by the prior alignment. In this formulation, we 
must update $\Tensor{\bigtriangleup}_G$ and $\Tensor{\bigtriangleup}_H$ prior to running the 
full TAME algorithm. The key idea is to remove triangles for which at least one of the end-points
has no homology suggested by the sequence similarity.
This allows us to focus on the most promising regions of the graph. The constrained TAME method
is presented in Algorithm~\ref{alg:TAME_sparse}. In this algorithm, we first compute a pair
of indicator vectors, 
$\Vector{w}_r$ and $\Vector{w}_c$ with size $|\Set{V}_G|$ and $|\Set{V}_H|$, 
respectively. Each element $i$ in $\Vector{w}_r$ indicates if vertex $v_i \in \Set{V}_G$ has at least 
one homolog among vertices of $\Graph{H}$ and, similarly, each element $i'$ in $\Vector{w}_c$ 
indicates if vertex $v_{i'} \in \Set{V}_H$ has at least one homolog among vertices of $\Graph{G}$ (determined
by the prior similarity). The 
``\textbf{.*}'' operator is the element-wise product of two tensors. Finally, we prune the triangle 
tensors by enforcing that all end-points of every triangle motif should have at least one homolog 
in the other graph. An equivalent way of understanding this algorithm is that we first remove
vertices from $\Graph{G}$ and $\Graph{H}$ that have no prior information indicating there is
a match in the other graph.

\begin{algorithm}
\centering
\begin{algorithmic}[1]
\REQUIRE Triangle tensors $\Tensor{\bigtriangleup}_G, \Tensor{\bigtriangleup}_H$;   Sequence similarities $\Vector{w}$; Shift parameter $\beta$
\ENSURE The final set of aligned node pairs $\langle \Vector{m_i}, \Vector{m_i'} \rangle$
\STATE $\Matrix{W} = \text{\textbf{unvec}}(\Vector{w})$
\STATE $\Vector{w}_G = \text{indicator vector for rows of $\Matrix{W}$ with non-zeros}$
\STATE $\Vector{w}_H = \text{indicator vector for cols of $\Matrix{W}$ with non-zeros}$
\STATE $\Tensor{W}_G = \Vector{w}_G \otimes \Vector{w}_G \otimes \Vector{w}_G$
\STATE $\Tensor{W}_H = \Vector{w}_H \otimes \Vector{w}_H \otimes \Vector{w}_H$
\STATE $\Tensor{\bigtriangleup}_G^{(\text{contrained})} = \Tensor{\bigtriangleup}_G .* \Tensor{W}_G$
\STATE $\Tensor{\bigtriangleup}_H^{(\text{contrained})} = \Tensor{\bigtriangleup}_H .* \Tensor{W}_H$
\STATE $\Matrix{X} = \textbf{TAME}(\Tensor{\bigtriangleup}_G^{(\text{contrained})}, \Tensor{\bigtriangleup}_H^{(\text{contrained})}, \Vector{w}, \beta)$
\end{algorithmic}
\caption{The contrained Triangular AlignMEnt (cTAME) algorithm}
\label{alg:TAME_sparse}
\end{algorithm}

This algorithm has the side-effect of reducing the total number of triangles (see Table~\ref{table:netStats}), 
resulting in a faster execution time. In many cases, 
it also outperforms the full version of TAME in terms of alignment quality by focusing the search
in more promising regions. 
We discuss the pros and cons of each of these methods in Section~\ref{sec:results}

\subsection{Post-processing algorithm}

The result matrix $\Matrix{X}$ returned by both TAME and cTAME is a heuristic for the integer problem~\eqref{eq:TAME} since it is real-valued and it does not enforce a one-to-one constraint during its
iterations. We propose a post-processing step, presented in Algorithm~\ref{alg:postProc}, that aims to maximize the objective~\eqref{eq:TAME} in the process of generating an integer solution from $\Matrix{X}$. The overarching idea of the post-processing algorithm is to examine small regions around each matched pair to find a local swap that either enhances topological similarity or preserves topological similarity and improves sequence similarity. These principles are similar to PISwap~\cite{PISWAP} and GHOST~\cite{GHOST} in that we swap matches using a greedy approach
to enhance the overall quality of the alignment.  Given a matching $M$, we define the \emph{sequence similarity} as $\sum_{ii' \in M} w(i,i')$ and the \emph{topological similarity} as $\sum_{ii', jj', kk' \in M} \Delta_G(i,j,k) \Delta_H(i',j',k') [1+\max(w(jj') + w(kk'), w(jk') + w(kj'))]$. Intuitively, this can be seen as a weighted average of triangles incident on each aligned pair $ii'$, which encourages additional sequence similarity. These are the two components we evaluate each time we consider a local swap and use the rule above to accept the swap.\footnote{In our implementation, we use an efficient routine to evaluate how much each metric changes rather than recomputing from scratch.} 

The way we implement the iterative swapping procedure is as follows. We begin post-processing by using the Hungarian algorithm to identify a max-weight bipartite matching on the TAME matrix $\Matrix{X}$. Then, we consider a fixed number of rounds of swapping. At the start of each round, we build a list of unprocessed matches based on the \emph{current matching that exists} at that point in time. This list is examined in order of largest weighted degree of the matched vertices in the bipartite graph with matrix $\Matrix{X}$ in order to increase the likelihood of a swap.  In each examination step, we consider a match $ii'$ and search for possible swaps within its local neighborhood (defined next) that increase quality. We then evaluate each potential swap and accept it if it increases the topological similarity or preserves the topological similarity and increases sequence similarity. At the end, if we decided to swap $ii'$ and $jj'$, then we implement the swap in the matching \emph{immediately} and update $jj'$ to $ji'$ in the list of unprocessed matches. (There is no re-sorting and this may be effectively  ignored if $jj'$ was already processed, itself.) 

The local neighborhood searched for improved matches consists of a $b$-matching over the $\Matrix{X}$ from TAME and another $b$-matching over the sequence similarity scores.\footnote{Formally, a \emph{b-matching} is a generalization of a matching that enables each node to match to $b$ neighbors. For weighted graphs, we can define a maximum weight $b$-matching as the subset of edges with the maximum sum of weights, in which none of the vertices is adjacent to more than $b$ neighbors.} We use a half-approximation algorithm~\cite{bMatching} to solve the $b$-matching problem in linear time~\cite{Arif}. Thus, when examining a matched edge, the set of alternative candidates is the union of the neighbors in the $b$-matching of topological scores from $\Matrix{X}$, the neighbors of the $b$-matching of the sequence similarity scores, and the neighbors in the two protein interaction networks ($\Set{N}_\Graph{H}(i')$ and $\Set{N}_\Graph{G}(i)$).  These local alternative sets are called \emph{preferred sets}~\cite{PISWAP} and $\text{Pref}_{H}(i)$ and $\text{Pref}_{G}(i')$ represent alternative matches for $i$ in graph $\Graph{H}$ and $i'$ in graph $\Graph{G}$, respectively. The set of possible swaps for $ii'$ comprise any other pair $jj'$ where $j' \in \text{Pref}_H(i)$ and $j \in \text{Pref}_G(i')$ and $j$ is already matched to $j'$, in which case the swap is $ij'$ and $ji'$. We also consider a swap $ii'$ to $ij'$ if $j' \in \text{Pref}_H(i)$ and $j'$ is unmatched (and likewise, $ii'$ to $ji'$ if $j \in  \text{Pref}_G(i')$ and $j$ is unmatched).

\begin{algorithm}
\centering\footnotesize
\begin{algorithmic}[1]
\REQUIRE Output matrix of TAME/cTAME $\Matrix{X}$; Sequence similarity matrix $\Matrix{W}$; matching degree for topological scores $b_{topo}$; matching degree for sequence similarities $b_{seq}$
\ENSURE Final alignment $M$
\STATE Set the initial matching $M$ based on solving a max-weight matching problem in $\Matrix{X}$
\STATE Set $\Set{M}_C$ to the union of a $b_{\text{seq}}$-matching in the sequence similarity $\Matrix{W}$ and a $b_\text{topo}$-matching in the matrix $\Matrix{X}$
\FOR{a fixed number of iterations}
\STATE Sort matches in $\Set{M}$ based on $\delta_{ii'} = \sum_i x({ii'}) + \sum_{i'} x({ii'})$ and set the unprocessed list $\mathcal{L}$ to these matches
\FOR {each unprocessed $ii' \in \mathcal{L}$ in the sorted order}
\STATE set $ii'$ as processed 
\STATE Set Pref$_\Graph{H}(i)$ = \{j'; $ij' \in \Set{M}_C$ \textbf{or} $j' \in \Set{N}_\Graph{H}(i')$\}
\STATE Set Pref$_\Graph{G}(i')$ = \{j; $i'j \in \Set{M}_C$ \textbf{or} $j \in \Set{N}_\Graph{G}(i)$\}
\FOR {each swap $S$ of the match $i$, $i'$ with another matched pair $j$,$j'$ in the preferred sets or any unmatched vertex in the preferred sets }
\STATE Check if $M$ with swap $S$ results in a higher topological similarity and accept it if it does, also accept if the topological score is equivalent, but the sequence score is higher. (These are defined in the text)
\ENDFOR
\STATE If the final swap $S$ contains a matched pair $j,j'$, then update $j,j'$ to $j,i'$ in $\mathcal{L}$, but do not resort.
\ENDFOR
\ENDFOR
\RETURN $\Set{M}$
\end{algorithmic}
\caption{Post-processing algorithm}
\label{alg:postProc}
\end{algorithm}

\section{Results and Discussion}
\label{sec:results}

\subsection{Datasets}

\subsubsection{Synthetic Datasets and Random Networks}

NAPAbench~\cite{NAPAbench}, is a family of random graphs that has been proposed for 
evaluating network alignment methods on synthetic datasets. This dataset contains both pairs of 
networks for evaluating pairwise alignment methods, as well as groups of networks for testing 
multiple alignment algorithms. There are three random graph generation models 
employed by NAPAbench: (i) duplication-mutation-complementation (DMC), (ii) duplication with 
random mutation (DMR), and (iii) crystal growth (CG). These random 
networks mimic key properties of biological graphs, including their network topology and modular structure.
We focus on the crystal growth (CG) dataset, which is based on a model that better fits features of real PPI networks, 
including their characteristic age distribution~\cite{Kim2008}.
This dataset contains 10 pairs of graphs, for which the known orthology and simulated 
sequence similarities between pairs of nodes are available. Each pair consists of a first graph 
$A$ with $3,000$ nodes and a second graph $B$ with $4,000$ nodes. These two networks share a 
common ancestor of size $2,000$ and have been evolved independently after that.  
Edge and triangle statistics for these graphs are summarized in 
Table~\ref{table:netStats_NAPA}

\begin{table}[!t]
\caption{Summary statistics for the NAPAbench dataset}
\label{table:netStats_NAPA}
\centering
\begin{tabular}{lccc}
\toprule
	&\# nodes	&Mean \# edges	&Mean \# triangles\\	\midrule
Graph A& 3,000 & 11,985	&11,362\\
Graph B& 4,000 & 15,985	&15,880 \\
\bottomrule
\end{tabular}
\end{table}

\subsubsection{Yeast Versus Human Interactome Dataset}
\label{sec:SeqSim}

Both yeast and human protein-protein 
interaction (PPI) networks were constructed from BioGRID database, version 3.2.103. All physical 
interactions, excluding self-loops and interspecies interactions, have been filtered and mapped to 
Entrez gene IDs. We used these interaction evidences to construct the adjacency matrix for both 
graphs. Edge and triangle  statistics for each network are presented in 
Table~\ref{table:netStats}.

\begin{table}[!t]
\caption{Summary statistics for yeast and human interactomes.}
\label{table:netStats}
\centering
\begin{tabular}{lccc}
\toprule
	&\# nodes	&\# edges	&\# triangles\\	\midrule
Human	&14,867	&126,593	&407,650\\
Yeast	&5,850	&79,458	&347,079\\ 
\addlinespace
constrained Human & 10,624 & 88,276 & 251,555\\
constrained Yeast & 5,482 & 73,739 & 289,893 \\
\bottomrule
\end{tabular}
\end{table}

We downloaded the protein sequences for the yeast and 
humans genes in FASTA format from Ensembl database, release 69. These datasets are based on the GRCh37 
and EF4 reference genomes, each of which contain 101,075 and 6,692 protein sequences for 
\textit{H. Sapiens} and \textit{S. Cerevisiae}, respectively. Each human gene in this dataset has, 
on average, around 4 protein isoforms. We identified and masked low-complexity regions in protein 
sequences using \textit{pseg} program~\cite{PSEG}. The \textit{ssearch36} tool, from 
\textit{FASTA}~\cite{FASTA} version 36, was then used to compute the local sequence alignment of 
the protein pairs using the Smith-Waterman algorithm~\cite{Smith1981}. We used this tool with the 
BLOSUM50 scoring matrix to compute sequence similarity of protein pairs in humans and yeast. All 
sequences with E-values less than or equal to 10 are recorded as possible matches, which results 
in a total of 664,769 hits between yeast and human proteins. For genes with multiple protein 
isoforms, coming from alternatively spliced variants of the same gene, we only record the most 
significant hit. The final dataset contains 162,981 pairs of similar protein-coding genes. After 
mapping these pairs to the human and yeast interactomes, we were able to find matches for 127,505 
node pairs in these networks.

\subsubsection{Tissue-specific gene expression dataset}

We downloaded the RNASeq dataset version 4.0 (\textit{dbGaP accession phs000424.v4.p1}) from the The Genotype-Tissue Expression (GTEx) project~\cite{Ardlie2015}. We processed each sample using the UPC~\cite{Piccolo2013}. For each gene, we recorded the alternatively spliced transcript with the highest activation probability in the sample. The final dataset contains the expression value of 23,243 genes across 2916 biological samples, which includes 30 different tissues/cell types.

\subsection{Evaluation Criteria}
\label{sec:eval_criteria}

For each alignment, we separately assess the topological quality of the alignment graph, as well as
the biological relevance of aligned nodes in the input graphs.  Additionally, for the NAPAbench synthetic dataset, we use known matches to compute the correctness of network alignment. Measures described here are adopted from a recent work by Meng \textit{et al.}~\cite{Meng2015} as general means to compare local versus global network alignments. Specifically, \textit{generalized $S^{3}$ (GS3)} and \textit{F-FP} are shown to be good surrogates for topological and biological quality of alignment with respect to edge conservation and gene ontology (GO) consistency.
We extend the concept of \textit{GS3} to the case of triangles, which we call \textit{triangular GS3 (tGS3)}. In addition, we introduce a new measure based on co-expression of genes to validate biological plausibility of network alignments.

Let $m(ii')$ be an indicator function for matching, that is, $m(ii')=1$ iff vertex $v_i \in V_G$ is matched to $v_{i'}$ in $V_H$. Furthermore, let $\hat{m}(ii')$ be an indicator of true alignments, which is one if vertex $i$ in $V_G$ is a true match for vertex $i'$ in $V_H$. Let $\Set{M}$ and $\Set{\hat{M}}$ be the set of aligned pairs and true alignments. Using this notation, we can formulate different performance measures are as follows:

\subsubsection{Node Correctness (NC)}

This measure is only defined for synthetic cases for which the true-alignment is known a priori. We can define precision and recall for each alignment as $P=\frac{|\Set{M} \cap \Set{\hat{M}}|}{|\Set{M}|}$ and $R=\frac{|\Set{M} \cap \Set{\hat{M}}|}{|\Set{\hat{M}}|}$, respectively. Then, the F-score of node correctness, denoted by \textit{F-NC}, can be computed as the harmonic mean of the precision and recall.

\subsubsection{Node Coverage (NCV)}
Let us define the ``\textbf{unique}'' operator that when applied to a list returns the unique elements of the list as a set. We will denote by $V_{\Graph{G}}^{(A)}$ the set of \textit{touched} vertices in $\Graph{G}$ under alignment $A$, which is computed as \textbf{unique}($[i]; \forall ii' \in \Set{M}$). $V_{\Graph{H}}^{(A)}$ can be defined similarly for graph $\Graph{H}$. Using this notation, \textit{NCV} can be computed as $\frac{|V_{\Graph{G}}^{(A)}| + |V_{\Graph{H}}^{(A)}|}{|V_{\Graph{G}}| + |V_{\Graph{H}}|}$.

\subsubsection{NCV-Generalized $S^{3}$ (GS3)} 
We can compute the total number of conserved edges from 
the edge-set of the alignment graph, i.e. $E_C = \{ ( ii', jj') \where (i, j) \in E_G, (i', 
j') \in E_H, \text{\textbf{ and }} m(ii') = m(jj') = 1\}$. Similarly, we can define the set of \textit{gapped edges} as $E_{\not C} = \{ ( ii', jj') \where (i, j) \in E_G, (i', j') \not\in E_H \text{\textbf{ or }} (i, j) \not\in E_G, (i', j') \in E_H , \text{\textbf{ and }} m(ii') = m(jj') = 1\}$. Then, \textit{GS3} measure is defined as $\frac{|E_C|}{|E_C|+|E_\not C|}$. For complete one-to-one alignments (each node in smaller graph is mapped to exactly one node in the larger graph), \textit{GS3} measure is equivalent to \textit{S3} measure. However, \textit{GS3} allows comparison with sparse as well as local alignment methods. One downside of \textit{GS3} is that it does not penalize for the size of alignment. For example, an alignment that only aligns one conserved edge will have perfect \textit{GS3}. To remedy this, we compute the geometric mean of \textit{NCV} and \textit{GS3}, which is called \textbf{NCV-GS3} measure.

\subsubsection{NCV-Triangular GS3 (tGS3)}
Similar to $GS3$, \textit{triangular GS3 (tGS3)} is defined on the basis of total number of 
conserved and gaped triangles. It can be represented with respect to the triangle-set of the alignment graph, 
$T_C = \{ (ii', jj', kk') \where (i, j, k) \in T_G, (i', j', k') \in T_H , \text{\textbf{ and }} m(ii') = m(jj') = m(kk') = 1\}$. We  define the set of gaped triangles as $T_{\not C} = \{ ( ii', jj', kk') \where (i, j, k) \in T_G, (i', j', k') \not\in T_H \text{\textbf{ or }} (i, j, k) \not\in T_G, (i', j', k') \in T_H , \text{\textbf{ and }} m(ii') = m(jj') = m(kk') = 1\}$. Triangular GS3 (tGS3) is defined as $\frac{|T_C|}{|T_C| + |T_{\not C}|}$. Similar to \textit{GS3} measure, we have to adjust for the size of the alignment. We define \textbf{NCV-tGS3} measure as the geometric mean of \textit{NCV} and \textit{tGS3}.

\subsubsection{Prediction accuracy of gene ontology (GO) terms}

When the true alignment is not known, we cannot use node correctness to directly assess the 
quality of aligned pairs. Instead, we need to use other measures as proxies for potential ortholog 
pairs. Here, we use Gene Ontology (GO)~\cite{Ashburner2000} to evaluate matches. 
To avoid terms that are predicted based on the sequence similarity, we only include the set of experimental annotations, namely terms with evidence codes \texttt{\textit{EXP}}, \texttt{\textit{IDA}}, \texttt{\textit{IPI}}, \texttt{\textit{IMP}}, 
\texttt{\textit{IGI}}, and \texttt{\textit{IEP}}. The final dataset includes $38,880$ annotations for yeast spanning $5,060$ genes, and 
$158,429$ annotations for $11,235$ human genes. Using these annotations, Meng \textit{et al.} suggested a procedure for masking true GO terms for gene pairs and predicting them using network alignments. Using predictions that match known alignments, we can define precision and recall for the function prediction (\textit{P-PF} and \textit{R-PF}), and finally compute the F-score of the prediction (\textit{F-PF}).

\subsubsection{Gene expression consistency}
Coordinated expression of genes have been used extensively to define gene co-expression networks (GCN). The idea behind it is that genes are more likely to be functionally related if they are expressed similarly in different contexts. We adopt this point of view and define co-expression of all gene pairs in human using tissue-specific expression profiles from the GTEx project. Co-expression of each gene pair is computed as the Pearson's correlation of their expression profile across different tissues/cell types. We hypothesize that genes incident to conserved edges are more likely to be functionally related, and as such, are more likely to have high co-expression score. To measure this, we define a background distribution for the co-expression of all edges in the human interactome and compare it with the co-expression distribution of conserved edges using different methods. We use one-sided Wilcoxon rank sum test to assess whether the median of conserved edges significantly differ from the median of background distribution (all physical edges).

\begin{figure*}[!t]
\centering
\hfil%
\subfigure[Node correctness\label{fig:NAPA_NC}]{\includegraphics[width=.33\linewidth]{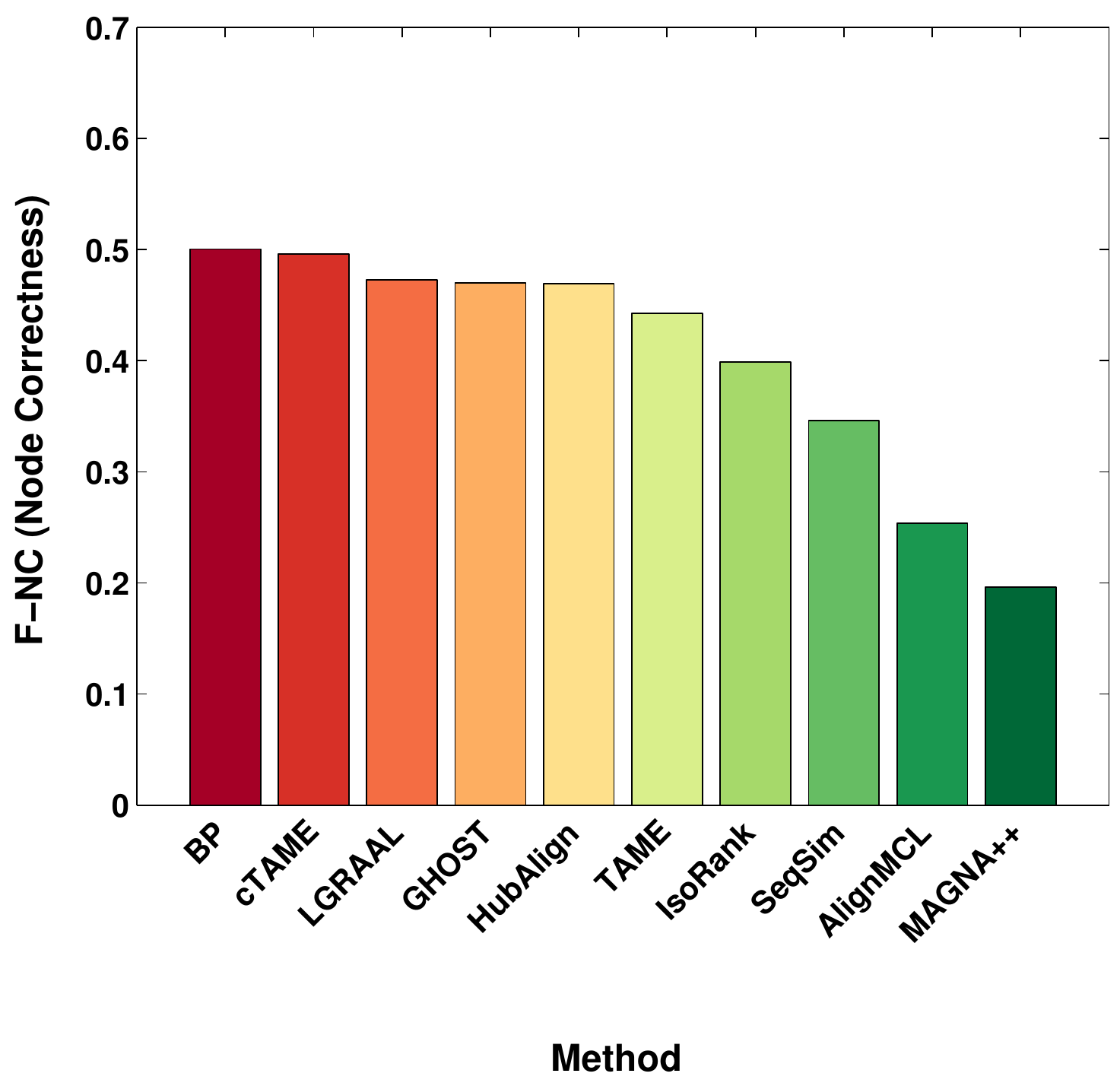}}%
\hfil%
 \subfigure[GS3 (\# edges)\label{fig:NAPA_NCV_GS3}]{\includegraphics[width=.33\linewidth]{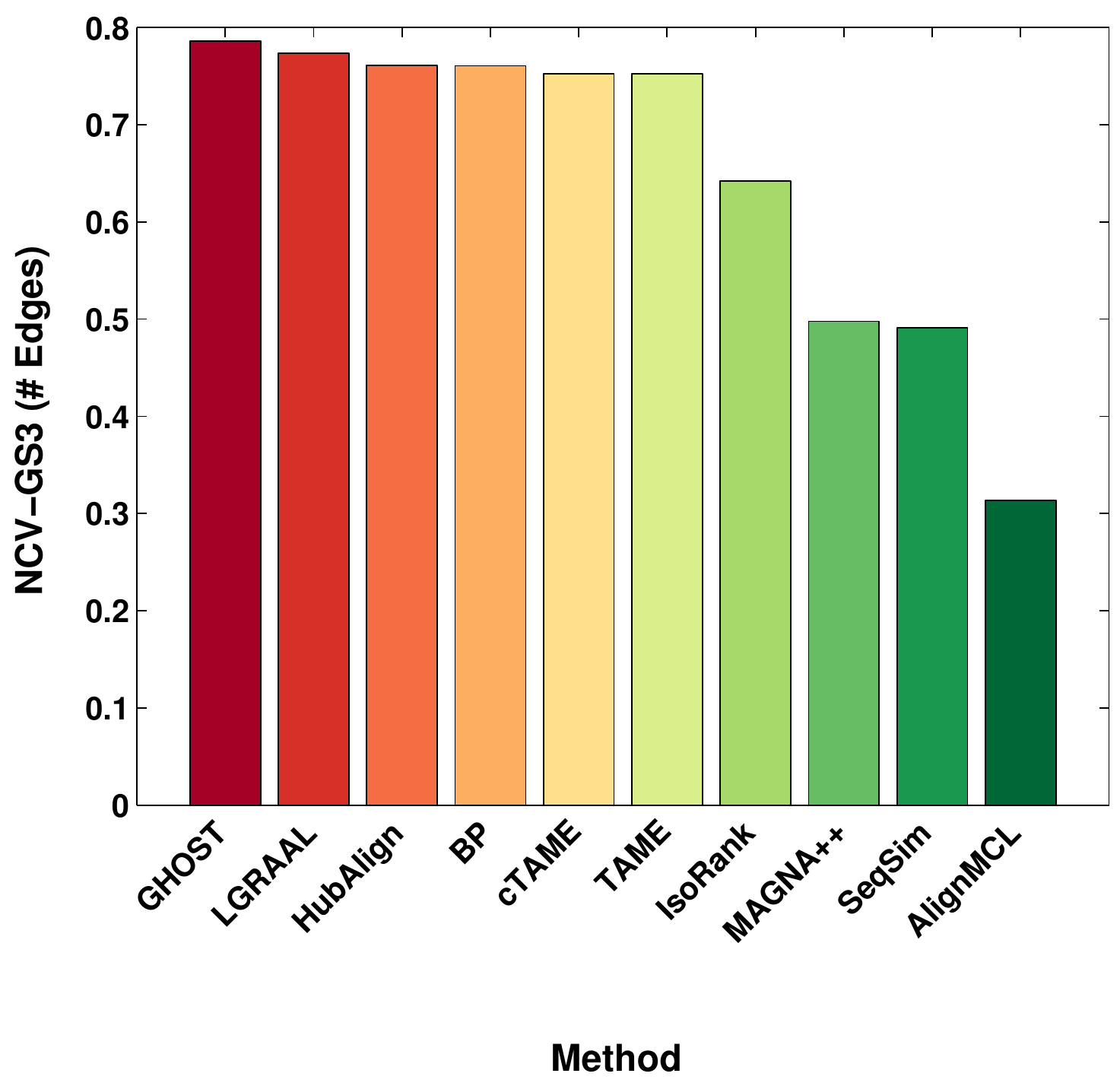}}%
 \hfil
 \subfigure[tGS3 (\# triangles)\label{fig:NAPA_NCV_tGS3}]{\includegraphics[width=.33\linewidth]{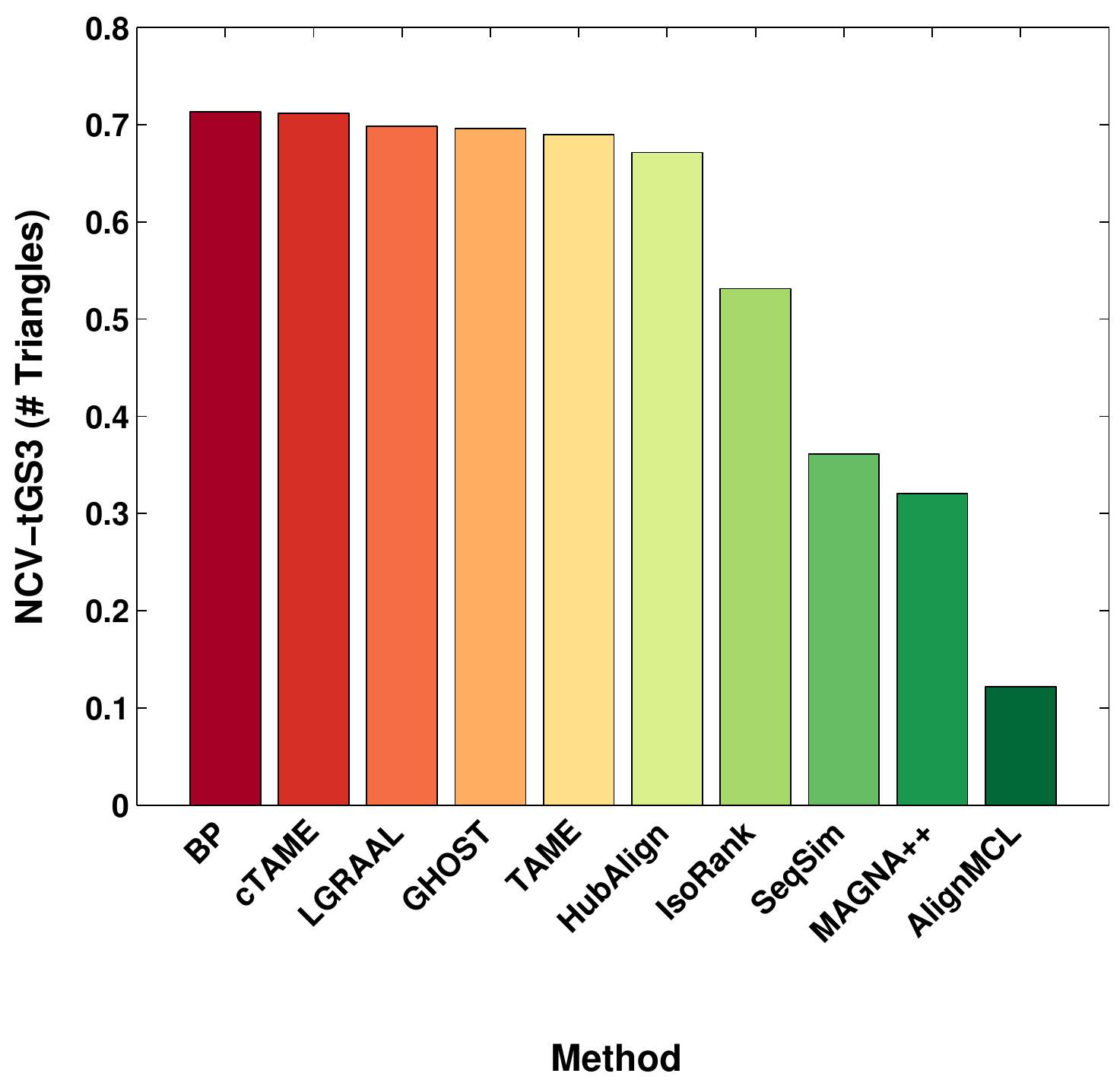}}%
 \hfil
 \caption{Comparison of alignment quality on NAPAbench synthetic dataset based on  the mean quality from 10 networks.}
 \label{fig:NAPA_validation}
\end{figure*}

\subsection{Experimental setting}
For methods that have an $\alpha$ parameter to balance topological/biological quality of alignments, we try three different values ($\alpha =\{ 0.15, 0.5, 0.85\}$) to find the optimal configuration. These points span a range between low and high topological influence (and high to low sequence influence).  To identify the optimal parameter, we first rank \textit{GS3, tGS3} and \textit{F-PF} (or \textit{F-NC} for NAPAbench dataset) scores for each method independently. Then, we choose the parameter that has highest average of ranks among these three measures. The goal is to find a \textit{unique} alignment that represents the \textit{best} quality both in terms of biology and topology. 
Table~\ref{table:params} summarizes the final set of parameters used in this study. In this table, \textbf{SeqSim} is simply the result of maximum weight matching (MWM) applied to the sequence similarity matrix.

\begin{table}[!h]
\centering
\begin{tabular}{ll}
 AlignMCL & No parameter\\ \addlinespace
 BP & $\alpha=3,\beta=17$ (NAPAbench), $\alpha=\beta=1$  (Sc\_vs\_Hs)\\ \addlinespace
 cTAME & main iterations=3, postprocessing iterations=3\\ &$\beta$=0 (NAPAbench), 1 (Sc\_vs\_Hs)\\ &$b_{topo}=200$, $b_{seq}=50$ \\ \addlinespace
 GHOST & $\alpha=0.5$ (Sc\_vs\_Hs)\\ & $\beta=1.0$ \\ & ratio=8.0 \\ &searchiter=10\\ \addlinespace
 HubAlign & $\alpha$= 0.85 (NAPAbench, Sc\_vs\_Hs) \\ \addlinespace
 IsoRank &  $\alpha$=0.85 (NAPAbench, Sc\_vs\_Hs) \\ \addlinespace
 L-GRAAL&time=86400s \\ &  $\alpha$=0.85 (NAPAbench), 0.5 (Sc\_vs\_Hs) \\  \addlinespace
 MAGNA++&  $\alpha$=0.15 (NAPAbench), 0.85 (Sc\_vs\_Hs) \\ & measure=S3\\&population size=15K\\&generations=2K\\ \addlinespace 
 SeqSim & No parameter\\ \addlinespace
 TAME & main iterations=3, postprocessing iterations=3\\ &$\beta$=0.1 (NAPAbench), 10 (Sc\_vs\_Hs)\\  &$b_{topo}=200$, $b_{seq}=50$ \\ \addlinespace
 
\end{tabular}
\caption{Parameter choices for different methods}
\label{table:params}
\end{table}

To tune the shift parameter $\beta$ in TAME/cTAME, we run the algorithm using values of the shift 
parameter over a log-linear search space ($\beta \in \{ 0, 
10^{-3}, 10^{-2}, 10^{-1}, 1, 10^{1}, 10^{2}, 10^{3} \}$) and choose the maximum 
based on the number of aligned triangles. 
We only report the value of the shift parameter with the best performance. 
The constrained and full formulations of TAME have different number of nonzeros in
the product tensors. For this reason, we run the parameter tuning phase 
for each of them independently.
In case of a random graph ensemble in NAPAbench, we compute the optimal shift values for each 
pair of networks independently, and use a majority voting technique to identify the value 
that performs the best in majority of alignments. More recent methods such as the generalized eigenproblem adaptive power (GEAP)~\cite{KoMa14} have been proposed as an extension of \textit{SS-HOPM}
that can automatically identify an adaptive shift in each iteration using the Hessian matrix of tensor-vector product.

%
%

\subsection{NAPAbench evaluation}

We align each pair of networks (a total of ten) separately using the different alignment methods. 
In Figure~\ref{fig:NAPA_validation}, we summarize various measures for the alignment quality of different methods when applied to the NAPAbench dataset. Figure~\ref{fig:NAPA_NC} shows the F-score of the node correctness, which is a measure of true alignment accuracy for each method. Sparse aligners, namely \textbf{BP} and \textbf{cTAME}, have similar performance, which is better than
the rest of the methods. \textbf{L-GRAAL, GHOST}, and \textbf{HubAlign} also have similar performance, which is marginally worse than sparse aligners and better than \textbf{TAME}. This effect can be explained by noting that true orthologs in NAPAbench, by construction, have nonzero sequence similarities. As such, these scores are highly informative for the true alignments and limiting the search space to the subset of pairs with known sequence similarity significantly simplifies the problem. \textbf{TAME} is primarily driven by topology and only uses sequence similarities directly in the post-processing step. Figure~\ref{fig:NAPA_NCV_GS3} and Figure~\ref{fig:NAPA_NCV_tGS3} show  measures of edge and triangle conservation under alignment. Results of conserved triangles is highly congruent with the actual node correctness, to a higher degree than edge conservation. To quantitatively measure this agreement, we computed the Kendall's tau (a nonparametric rank correlation) and its corresponding \emph{p}-value between the ordering of node correctness scores and edge/triangle conservation. We observed a correlation of 0.6 (\emph{p}-val=$0.02$) and 0.91 (\emph{p}-val=$2.98 \times 10^{-5}$) between \textit{F-NC} and \textit{NCV-GS3}/\textit{NCV-tGS3}, respectively. This suggests that both of these measures are positively related to the node correctness; however, between the two measures, triangle conservation is more significantly associated.

\subsection{Alignment of human versus yeast interactomes}

To assess the performance of different alignment methods when applied to real networks, we 
ran experiments on the yeast and human interactomes.  Recall that we cannot compute node correctness in this
case, since true orthologs are unknown. Therefore, we use F-score of GO function prediction as a proxy. 

\begin{figure}[!h]
\includegraphics[width=0.9\columnwidth]{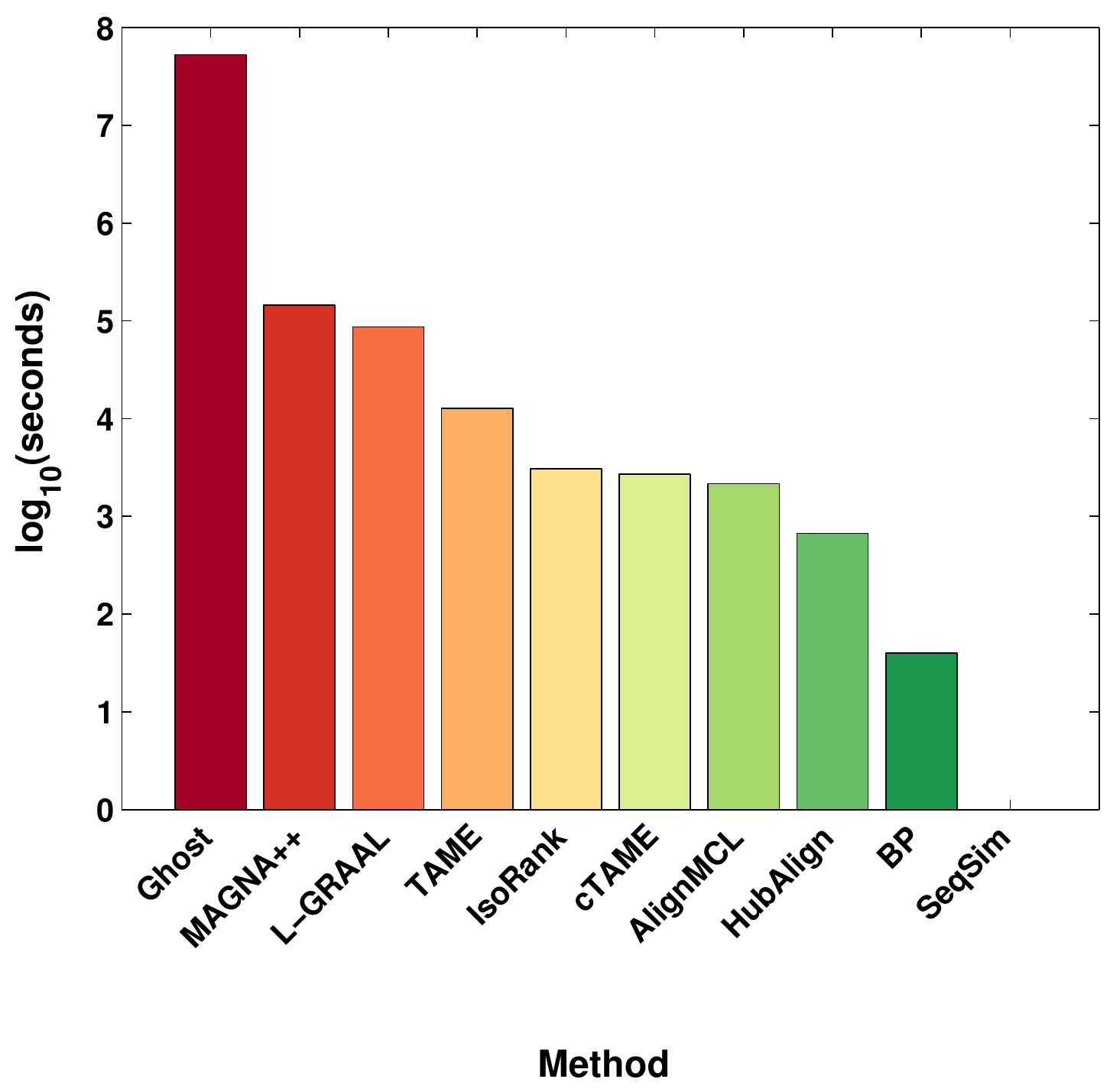}
\caption{Total amount of time taken by each alignment method}
\label{fig:timing}
\end{figure}

Various methods differ greatly in terms of total execution time. Figure~\ref{fig:timing} presents the running time of different methods when aligning yeast and human interactomes, reported as $log_{10}$ of the number of seconds for each method. For \textbf{SeqSim} method, it took less than a second to run, which we rounded up to 1s, the log of which is zero. At the other end of the spectrum, \textbf{GHOST} took 611 days of computational time on a single CPU to finish (we executed GHOST using 32 cores in parallel).

\begin{figure*}[!t]
\centering
\hfil%
\subfigure[F-score of GO function prediction\label{fig:F_PF}]{\includegraphics[width=.33\linewidth]{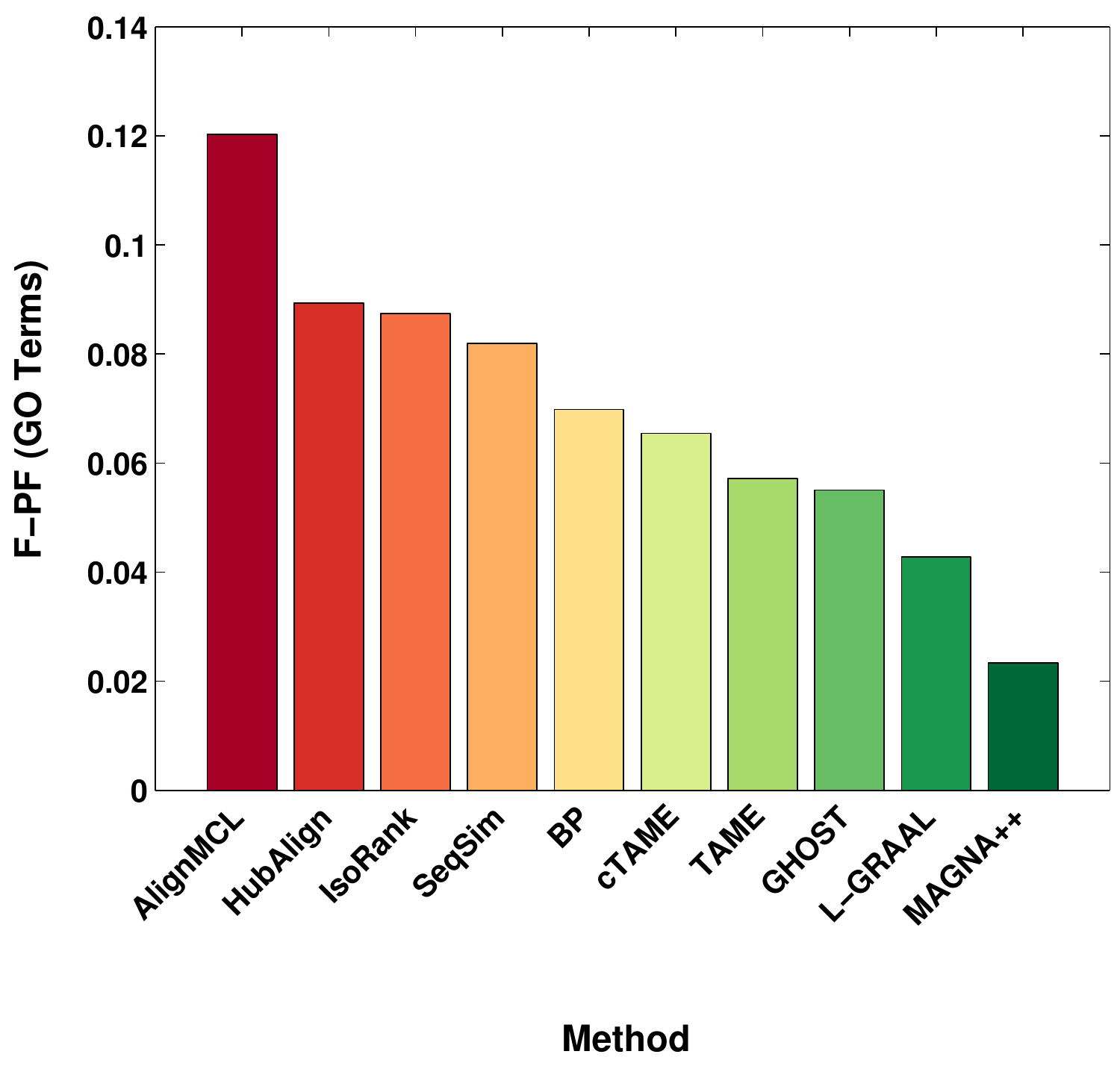}}%
\hfil%
 \subfigure[GS3 (\# edges)\label{fig:NCV_GS3}]{\includegraphics[width=.33\linewidth]{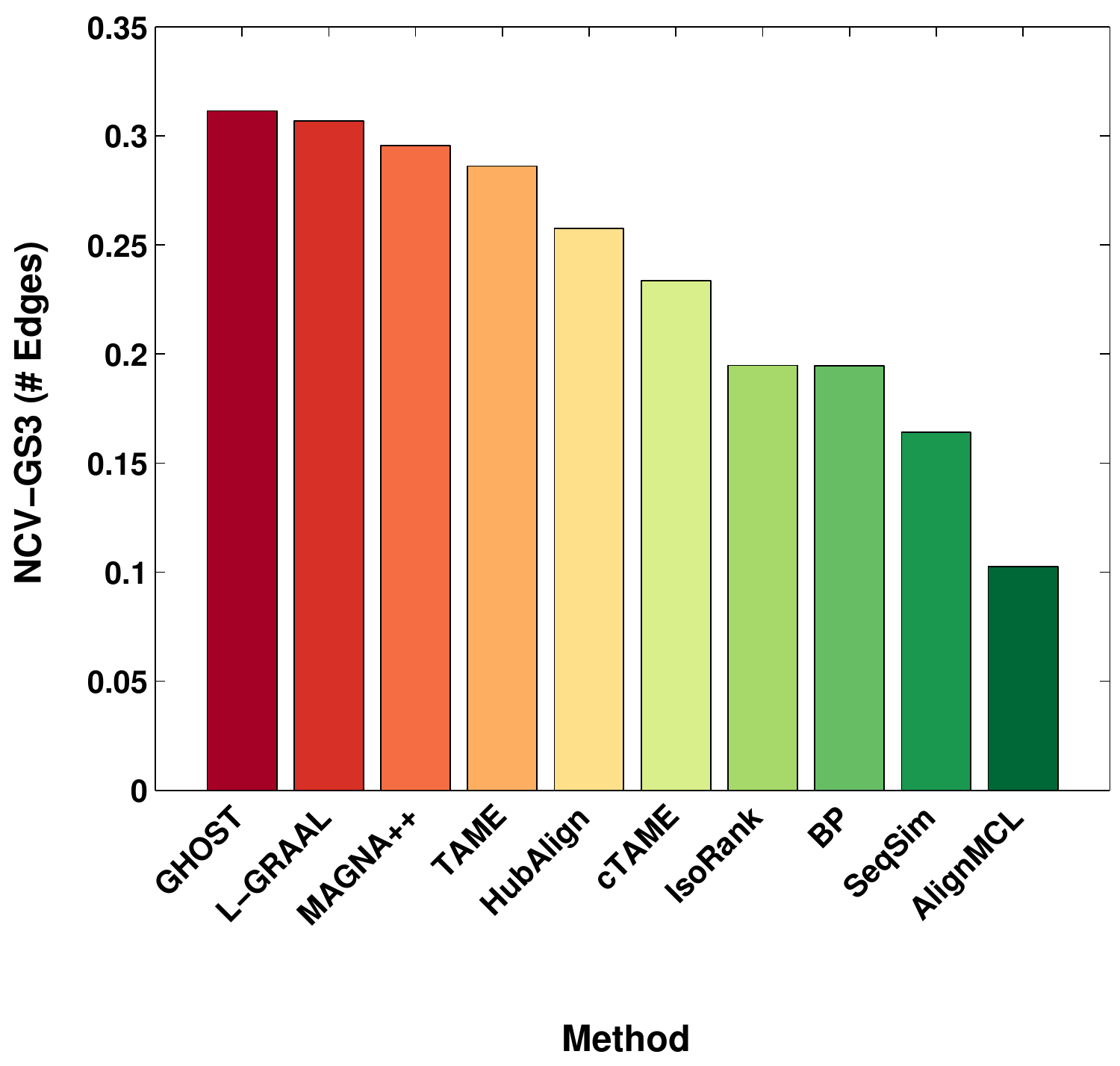}}%
 \hfil
 \subfigure[tGS3 (\# triangles)\label{fig:NCV_tGS3}]{\includegraphics[width=.33\linewidth]{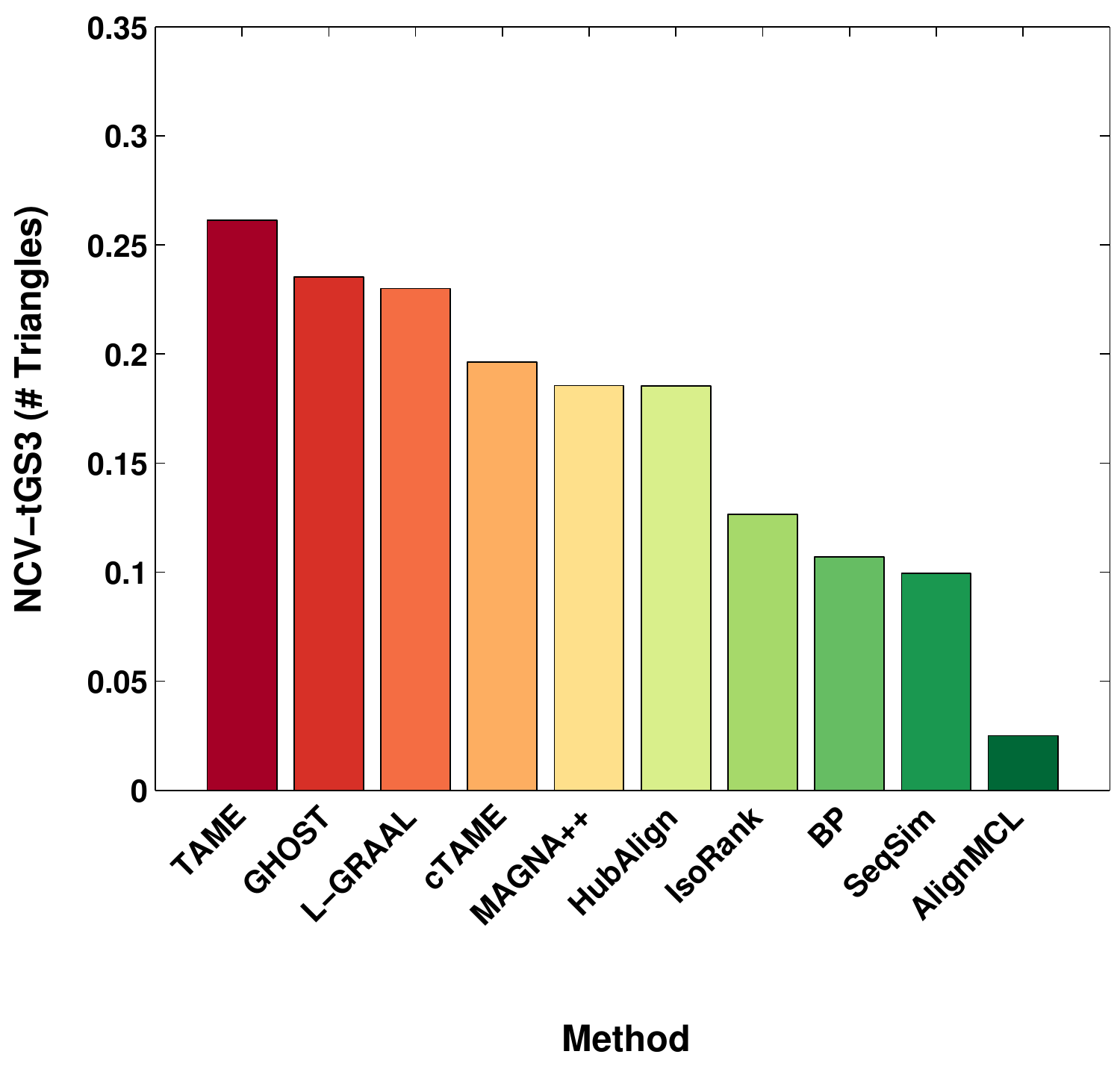}}%
 \hfil
 \caption{Comparison of alignment quality on yeast versus human dataset.}
 \label{fig:SCvsHS_validation}
\end{figure*}

In terms of alignment quality, Figure~\ref{fig:SCvsHS_validation} summarizes various measures computed for each alignment 
method. Figure~\ref{fig:F_PF} shows biological quality of the results, whereas Figures~\ref{fig:NCV_GS3}~and~\ref{fig:NCV_tGS3} illustrate edge and triangle conservation, respectively. Unlike NAPAbench, we observe a considerable difference between topological quality of alignments and their \textit{F-PF} scores. Our results, similar to Malod-Dognin \textit{et al.} \cite{L-GRAAL}, suggests that topological quality of alignments, both in terms of edges and triangles, negatively impacts the biological quality of alignments, measured as the prediction power for GO annotations. From biological point of view, we observed that \textbf{AlignMCL}, which is a local aligner, has the highest agreement with the GO annotations. This is consistent with the observation of Meng \textit{et al.} \cite{Meng2015} reporting that local aligners generally outperform global aligners in their prediction power for GO annotations. On the other hand, four methods with the best topological quality, namely \textbf{TAME, GHOST, L-GRAAL,} and \textbf{MAGNA++}, had the lowest F-score for predicting GO terms. 
In terms of edges, three methods that specifically optimize for conserved edges (\textbf{GHOST, L-GRAAL,} and \textbf{MAGNA++}) rank higher than \textbf{TAME}. On the other hand, in terms of triangles, \textbf{TAME} ranks the highest, followed by \textbf{GHOST} and \textbf{L-GRAAL}. It is notable here that the difference between the top-ranked method and runner up, in terms of edges, is $8.3\% (\frac{24,961 - 23,043}{23,043})$, whereas in case of triangles it is $18.6\% (\frac{76,403 - 64,433}{64,433})$. Moreover, note that \textit{NCV-GS3} and \textit{NCV-tGS3} are not monotonic in the number of edges/triangles. For example \textbf{MAGNA++} has 14,596 conserved edges while TAME has 20,569. However, \textbf{MAGNA++} has higher \textit{NCV-GS3} score.

Given the negative impact of edge/triangle conservation on the \textit{F-PF} scores, we aim to find which one of them is more detrimental to the quality of GO predictions. To this end, we compute the Kendall correlation between \textit{F-PF} and \textit{NCV-GS3} and \textit{NCV-tGS3}, individually. Both of the correlations are negative, confirming their negative impact. However, edge conservation scores  are significantly negatively correlated with \textit{F-PF} (\emph{p}-val = $1.6 \times 10^{-2}$), whereas the negative correlation of triangle conservation scores (\textit{NCV-tGS3}) and \textit{F-PF} is not significant (\emph{p}-val = $7.2 \times 10^{-2}$), at the significance threshold of 0.05. This suggests that methods that have higher number of conserved edges impact \textit{F-PF} more negatively than methods that have higher number of triangles. 

To further investigate whether these results are due to the nature of GO annotations, such as heavy bias towards gene pairs with high sequence similarity,  or due to the lack of biological signal, we propose a new biological measure. The main idea behind this measure is that proteins that have physical interaction are more likely to be functionally related. On the other hand, functionally related proteins are more likely to be similarly expressed in different contexts, such as different tissues/cell-types. We hypothesize that conserved edges are enriched with edges that are members of similar pathways/protein complexes and have lower false positive rate than the original network. Thus, distribution of co-expression scores between all gene pairs should have lower median than co-expression of interacting gene pairs, which in turn should have lower median than conserved edges under alignment. Using this hypothesis, we claim that a method recovers better alignment if conserved edges in the human interactome exhibit more significant difference from the background distribution of all edges.

\begin{figure}[!t]
\centering
\includegraphics[width=\columnwidth]{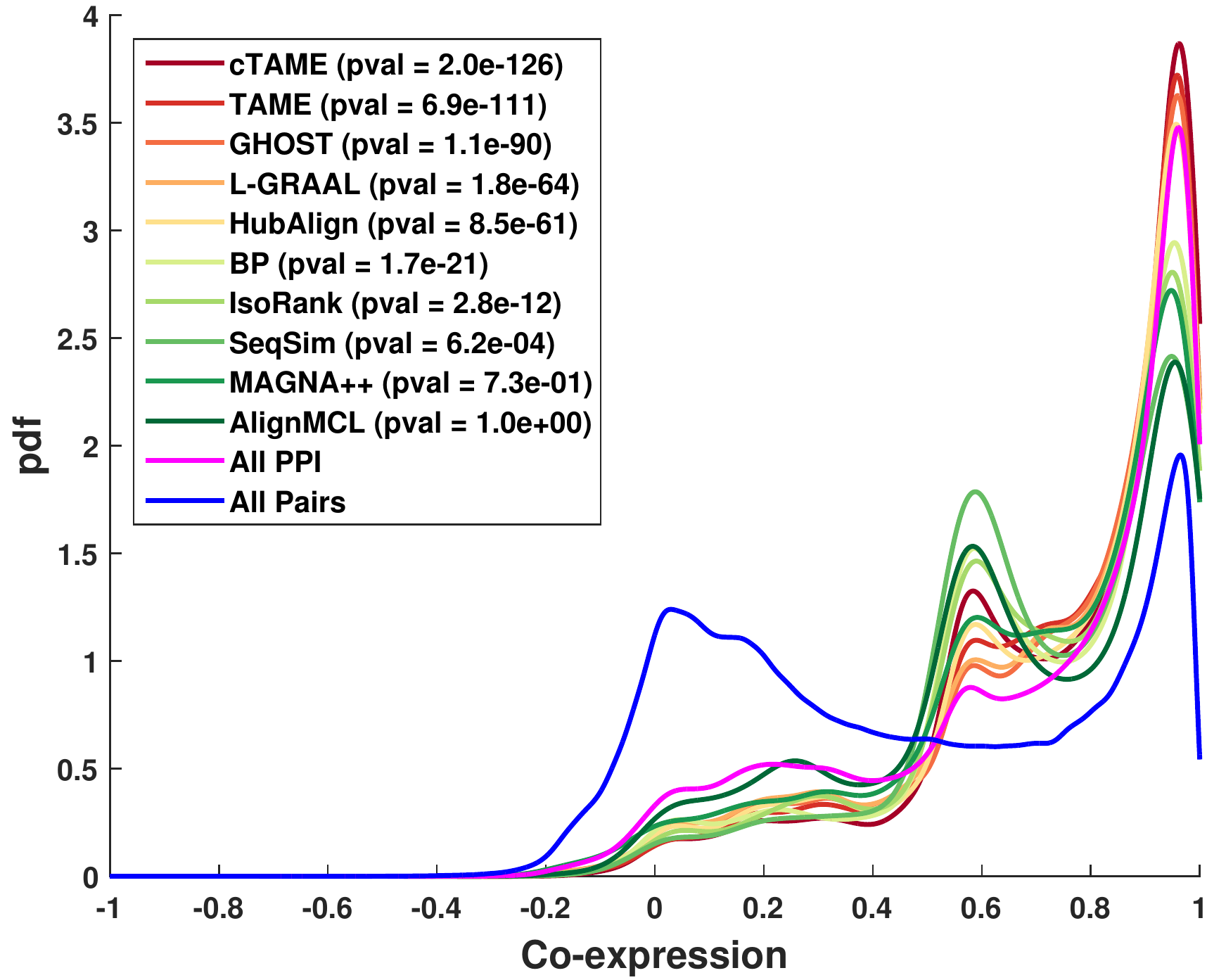}
\caption{Distribution of co-expression of gene pairs. Computed \emph{p}-value measure the significance of the median co-expression of conserved edges being larger than the background distribution of all physical edges in the human interactome}
\label{fig:CoExp}
\end{figure}

Figure~\ref{fig:CoExp} shows the distribution of co-expression values among different subsets of gene pairs in human interactome. First, we observe that the median of co-expression distribution for all gene pairs ($0.42$) is much smaller than the distribution for incident gene pairs on the physical edges ($0.78$). Next, we note that the left tail of co-expression distribution is heavier when considering all edges compared to only the subset of edges that are conserved in different methods.
For the subset of conserved edges, there are two main peaks in the distribution, one around 0.6 and the other around 0.9. We measured the significance of change in median between the background distribution of all edges and the subset of conserved edges in each method using the Wilcoxon rank sum test. Except \textbf{AlignMCL} and \textbf{MAGNA++}, all other methods show significant shift of median to the right (\emph{p}-value cutoff=0.05). All of these methods redistribute the heavy tail of PPI co-expression density to the peak of 0.6. However, among these methods, only \textbf{cTAME, TAME, GHOST, L-GRAAL} and \textbf{HubAlign} have denser (compared to the background) peaks around 0.9. \textbf{cTAME} and \textbf{TAME} show the most significant shift in the median among all other methods, with medians $0.84$ and   $0.82$, respectively. This suggests that the observed inconsistency between \textit{F-PF} measure and \textit{GS3}/\textit{tGS3} measures is not due to the lack of biological signal, but is attributed to the nature of GO terms.

%
%
%
%
%

\subsection{Behavior of TAME's iterations}
\begin{figure*}[!t]
\centering
\subfigure[Iteration 1]{\includegraphics[width=.19\textwidth]{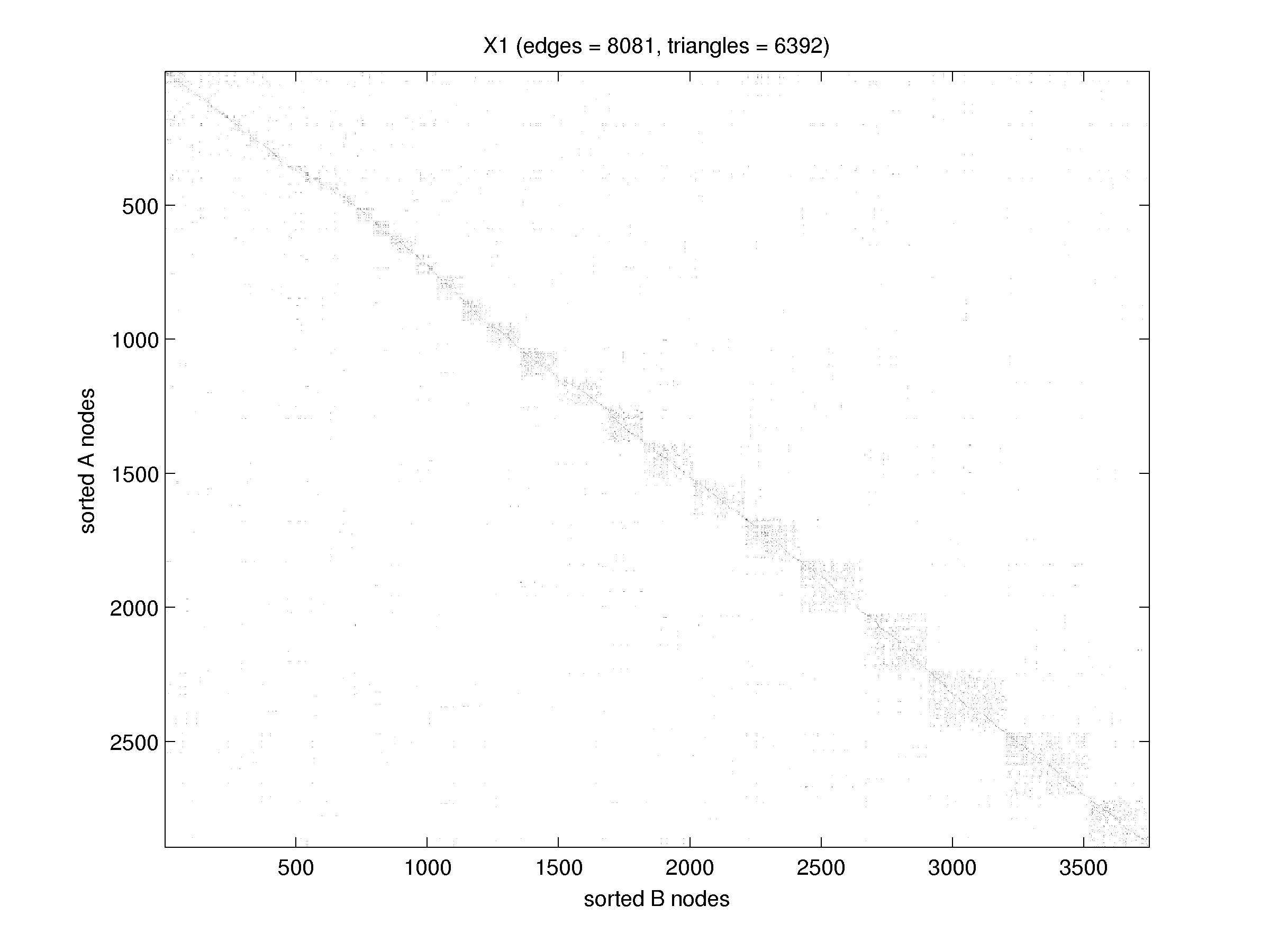}%
\label{fig:NAPA_X1}}
\hfil
\subfigure[Iteration 2]{\includegraphics[width=.19\textwidth]{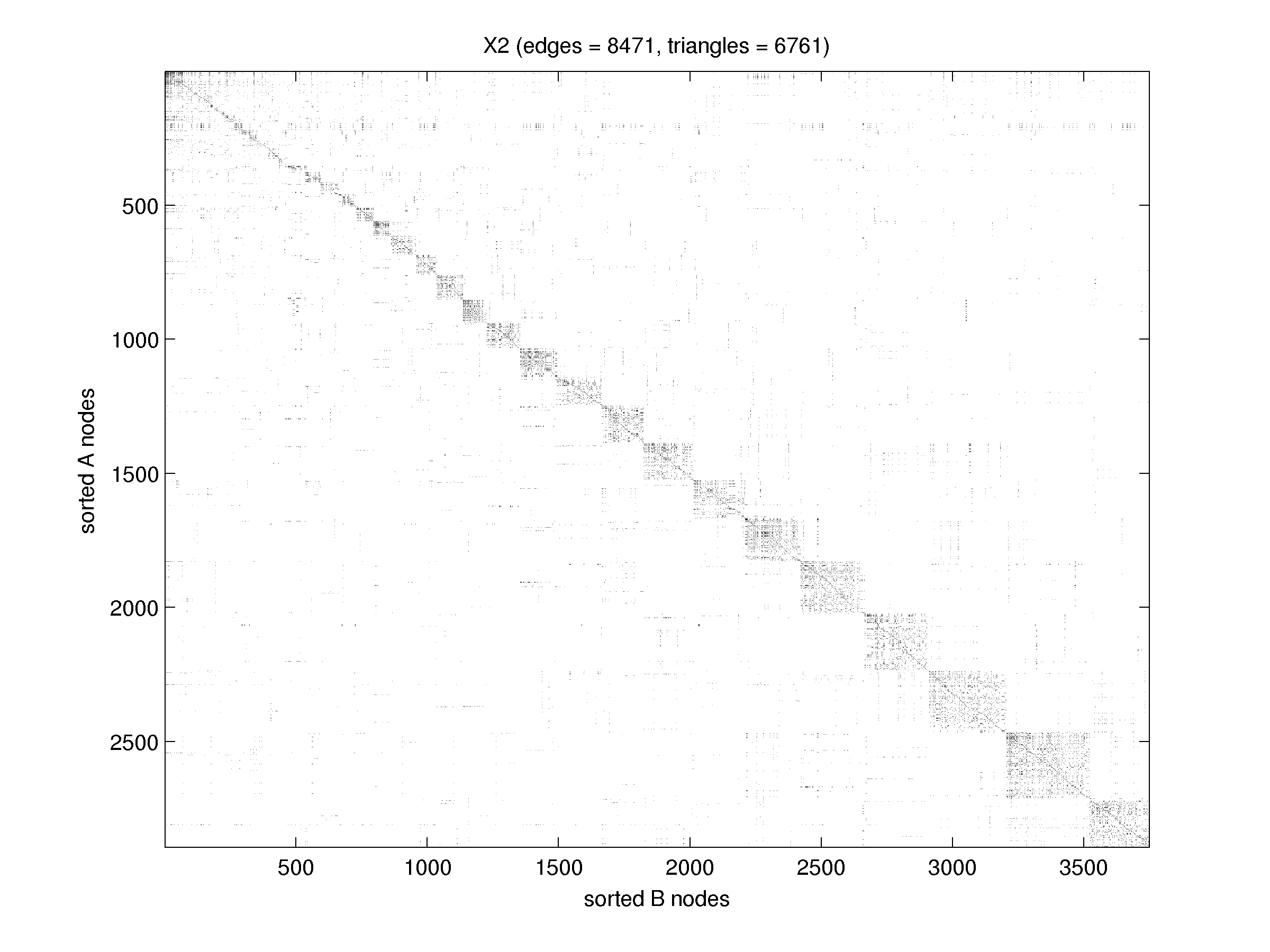}%
\label{fig:NAPA_X2}}
\hfil
\subfigure[Iteration 3]{\includegraphics[width=.19\textwidth]{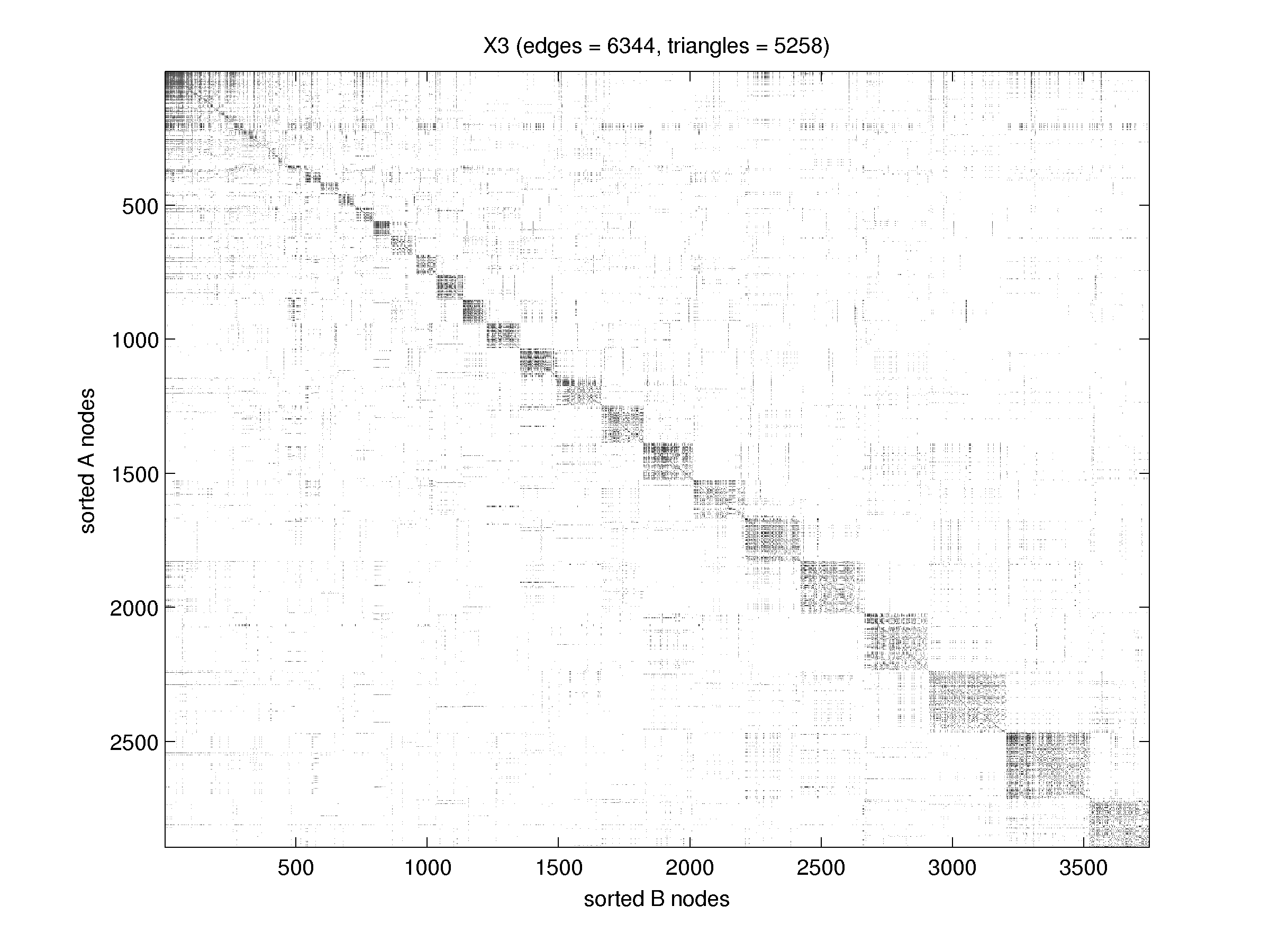}%
\label{fig:NAPA_X3}}
\hfil
\subfigure[Iteration 4]{\includegraphics[width=.19\textwidth]{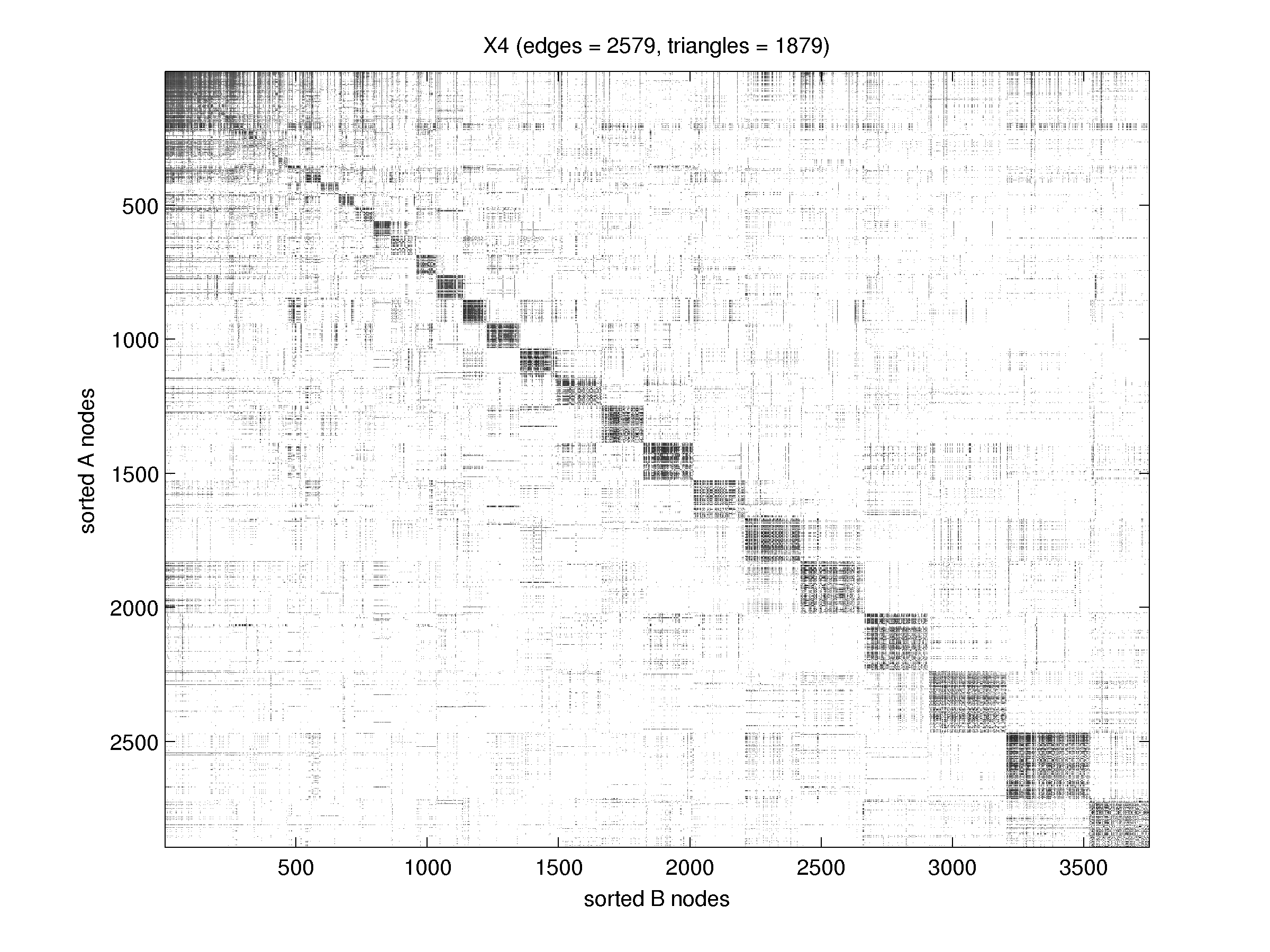}%
\label{fig:NAPA_X4}}
\hfil
\subfigure[Iteration 5]{\includegraphics[width=.19\textwidth]{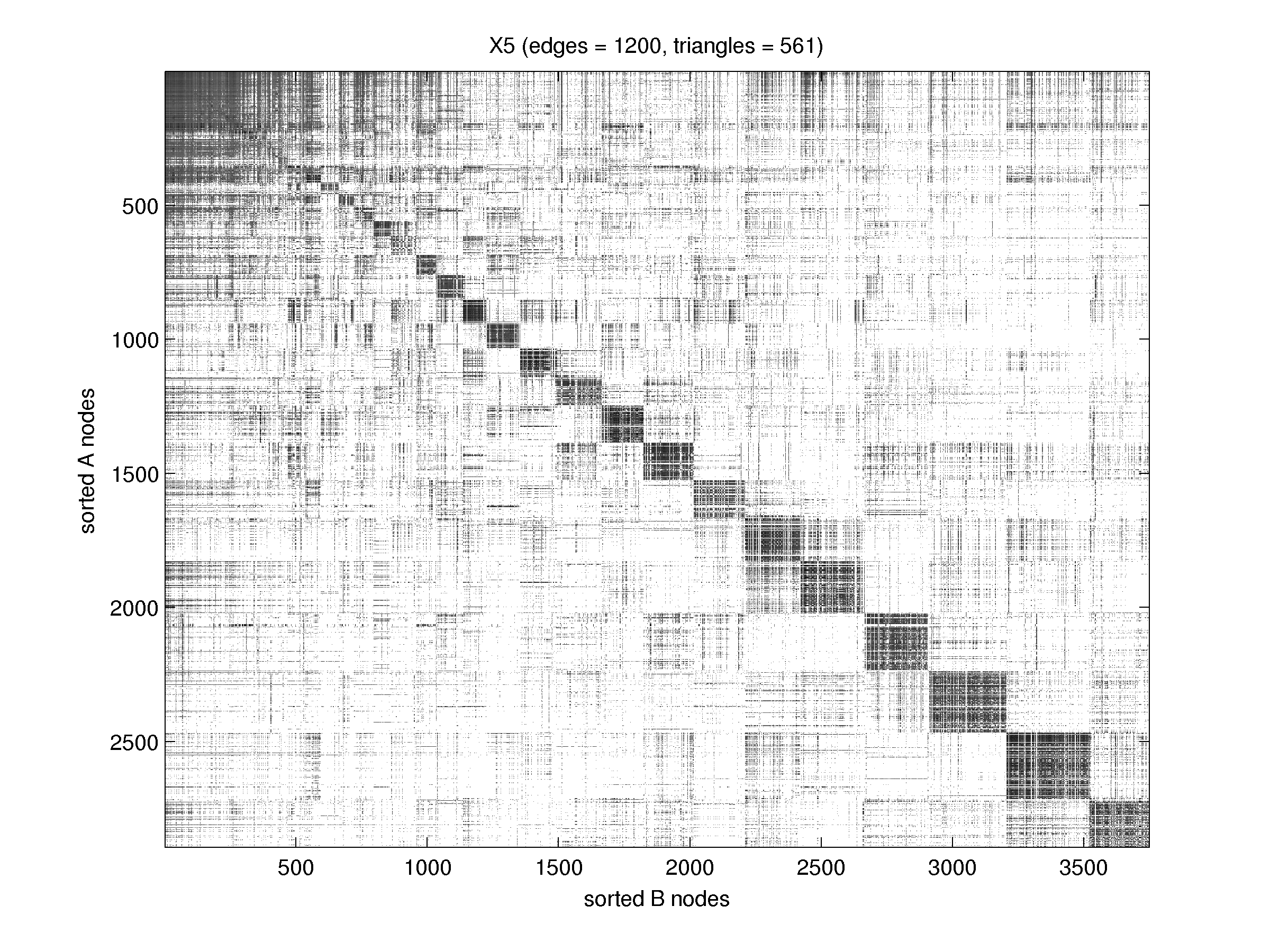}%
\label{fig:NAPA_X5}}

\subfigure[Iteration 6]{\includegraphics[width=.19\textwidth]{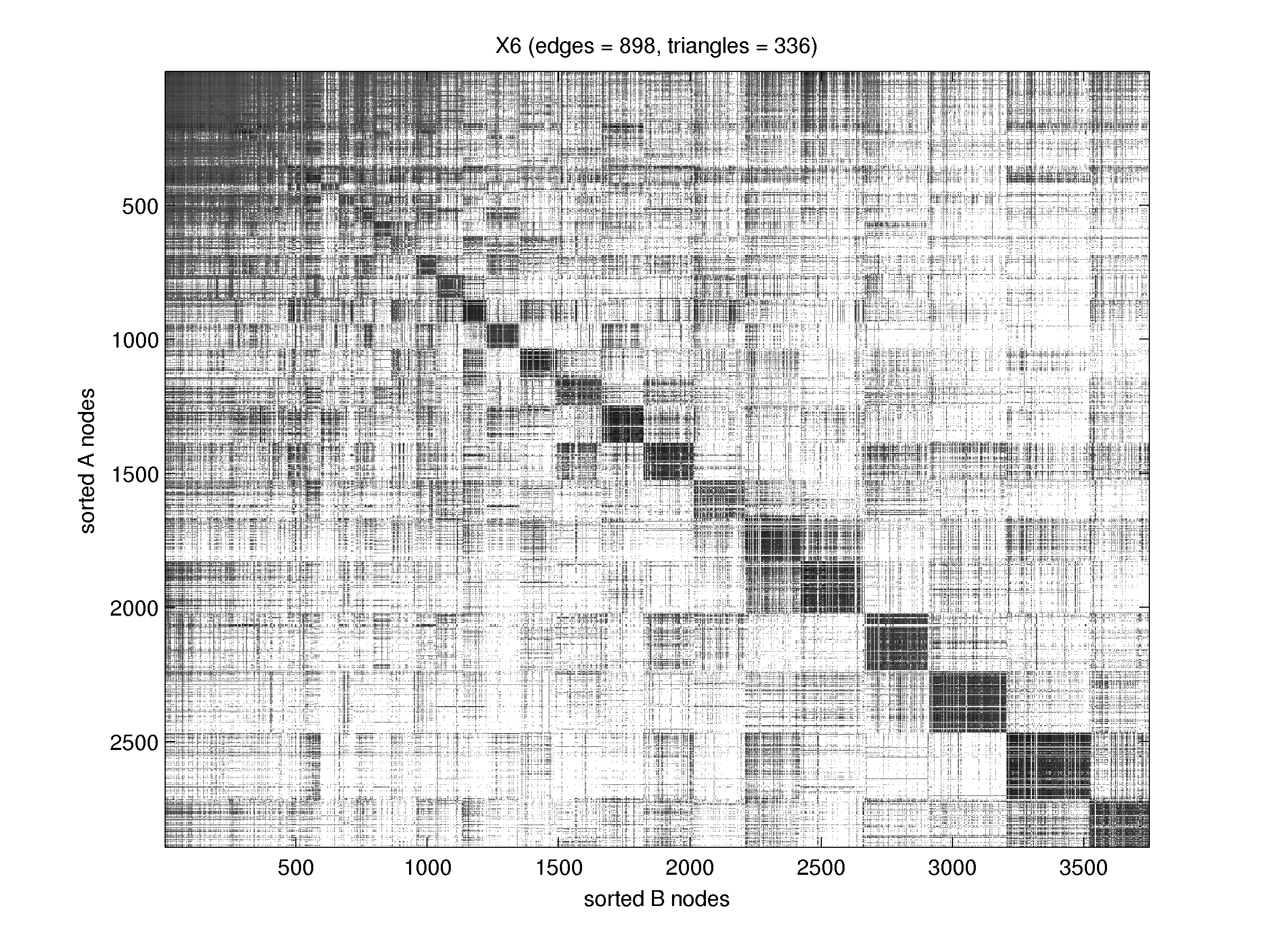}%
\label{fig:NAPA_X6}}
\hfil
\subfigure[Iteration 7]{\includegraphics[width=.19\textwidth]{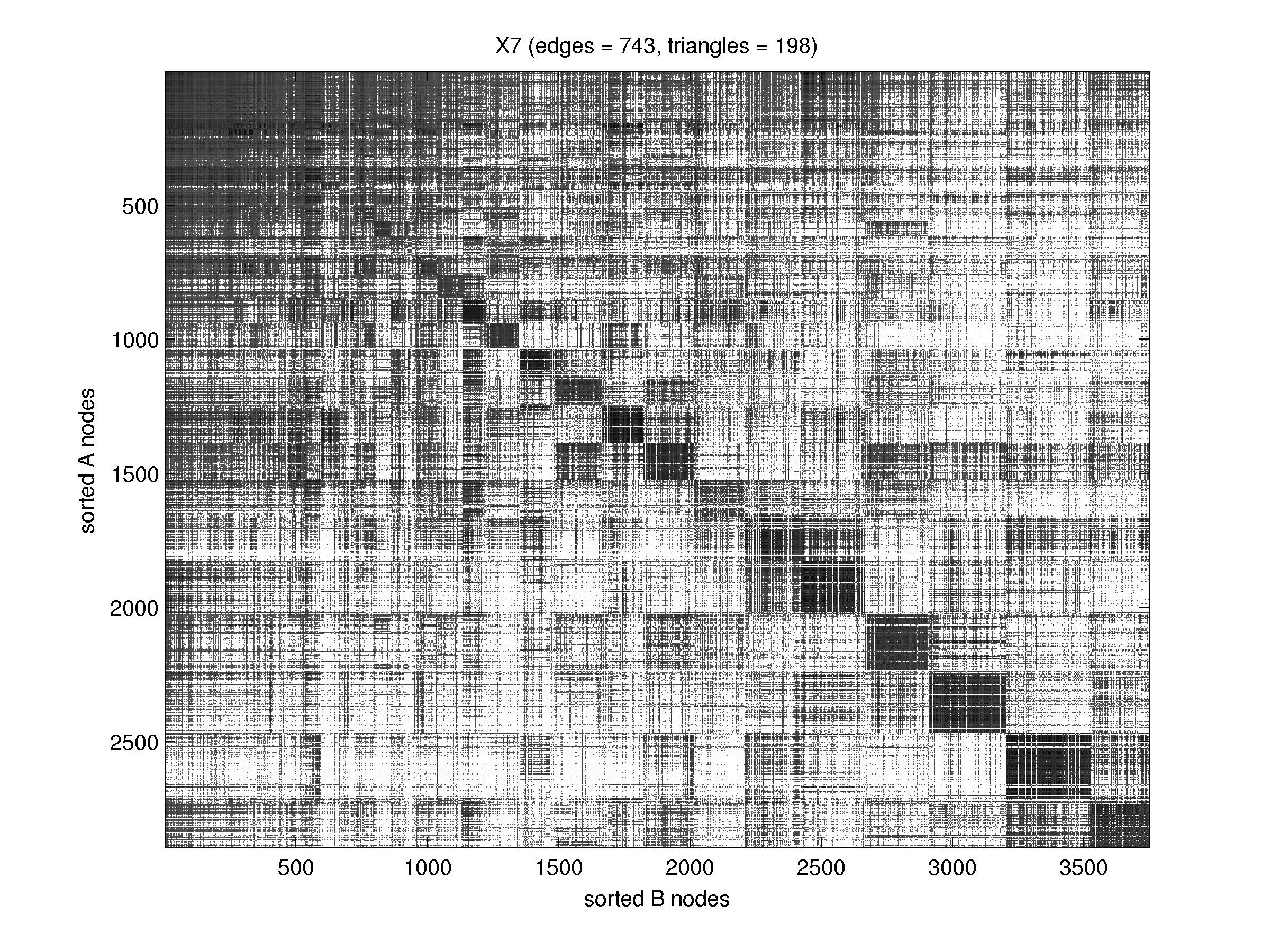}%
\label{fig:NAPA_X7}}
\hfil
\subfigure[Iteration 8]{\includegraphics[width=.19\textwidth]{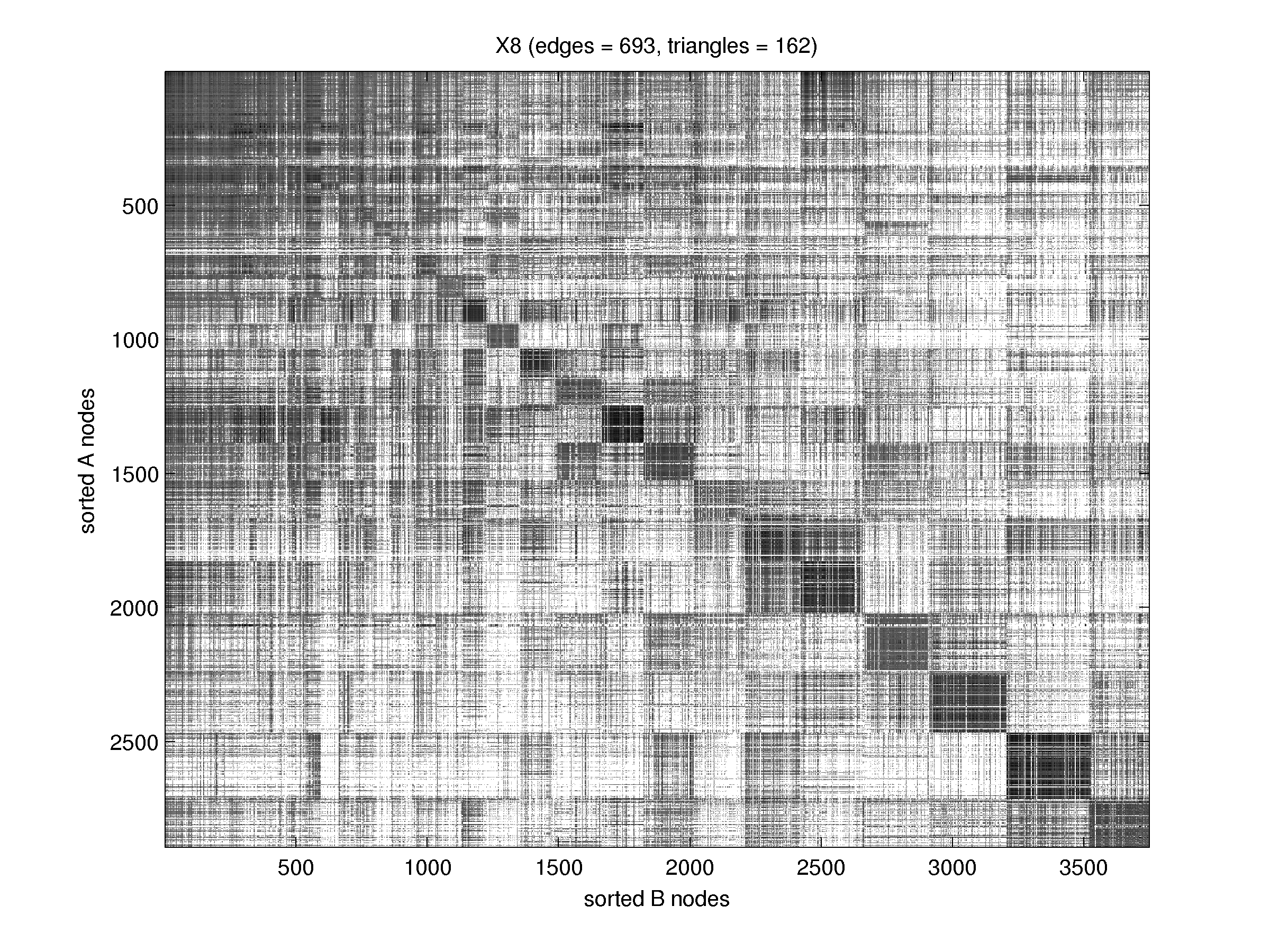}%
\label{fig:NAPA_X8}}
\hfil
\subfigure[Iteration 9]{\includegraphics[width=.19\textwidth]{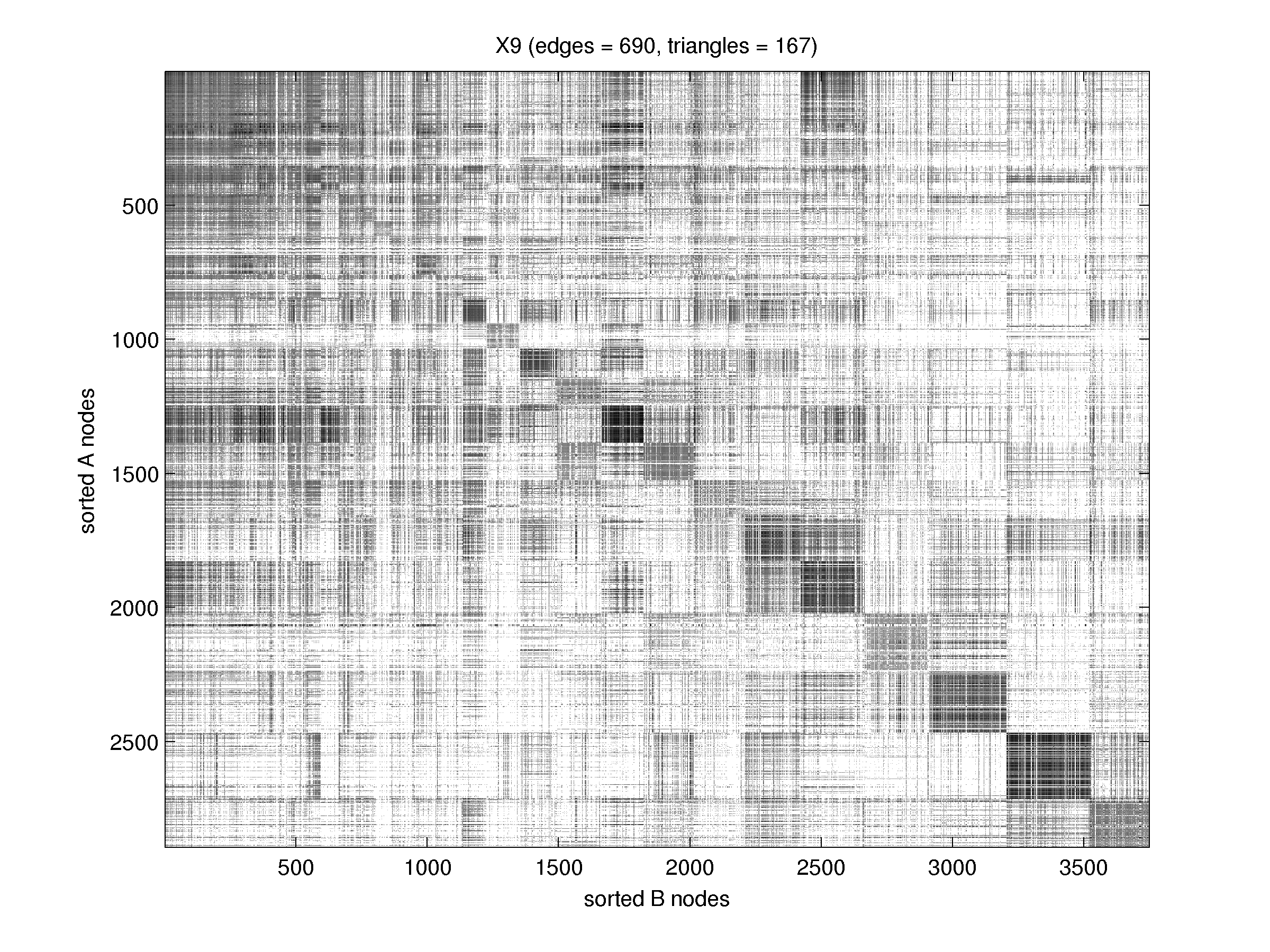}%
\label{fig:NAPA_X9}}
\hfil
\subfigure[Iteration 10]{\includegraphics[width=.19\textwidth]{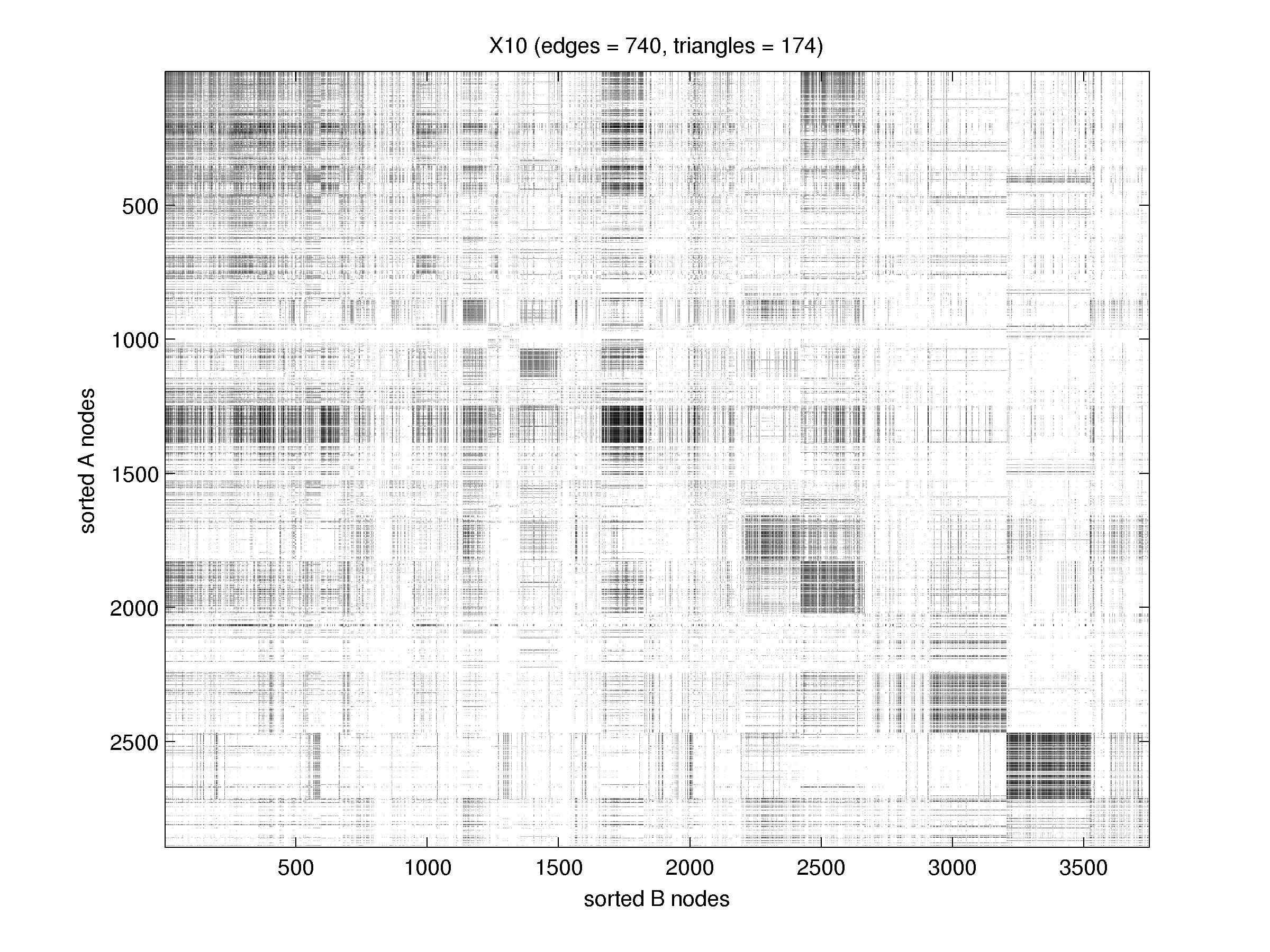}%
\label{fig:NAPA_X10}}

\subfigure[Iteration 11]{\includegraphics[width=.19\textwidth]{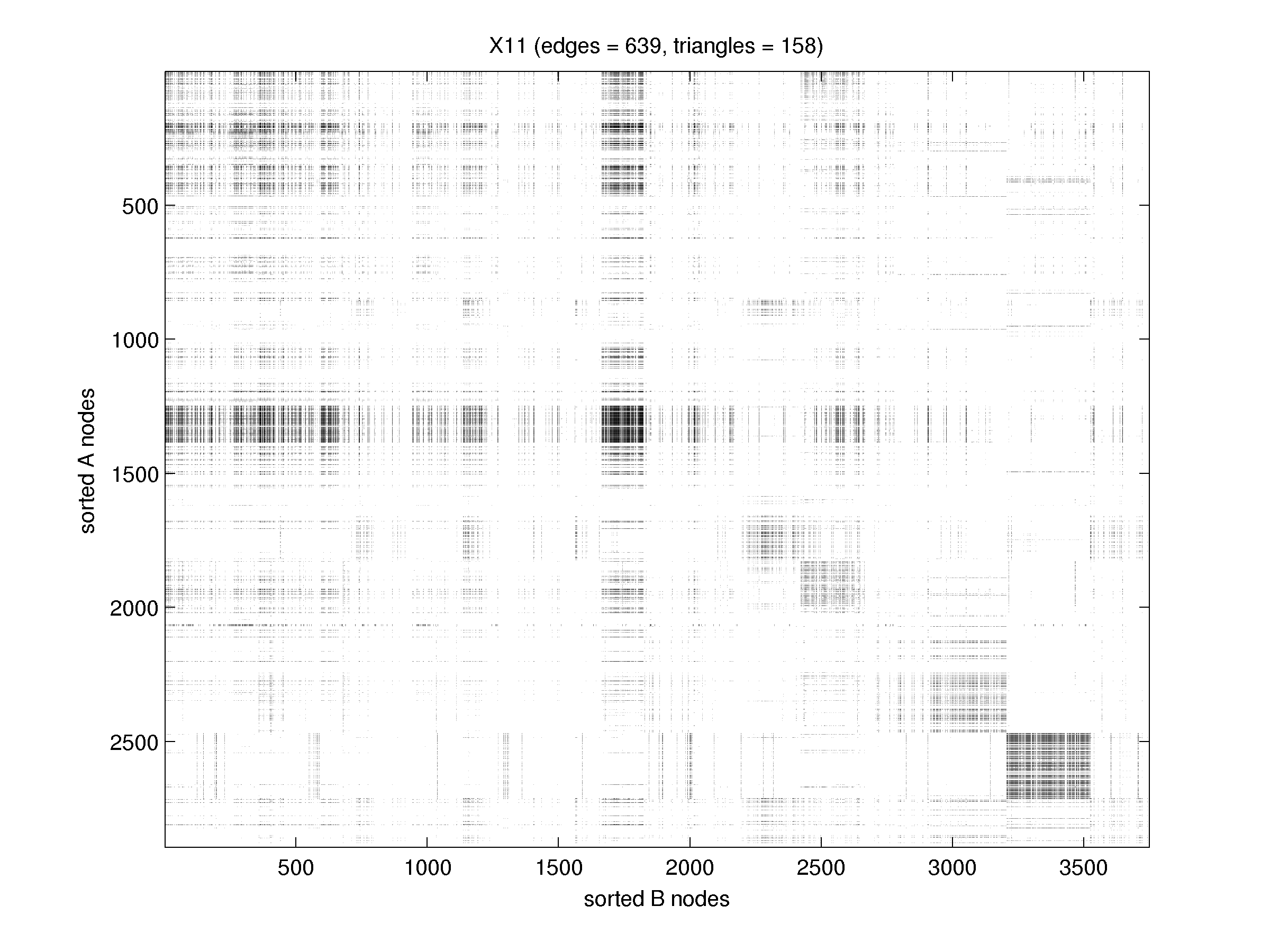}%
\label{fig:NAPA_X11}}
\hfil
\subfigure[Iteration 12]{\includegraphics[width=.19\textwidth]{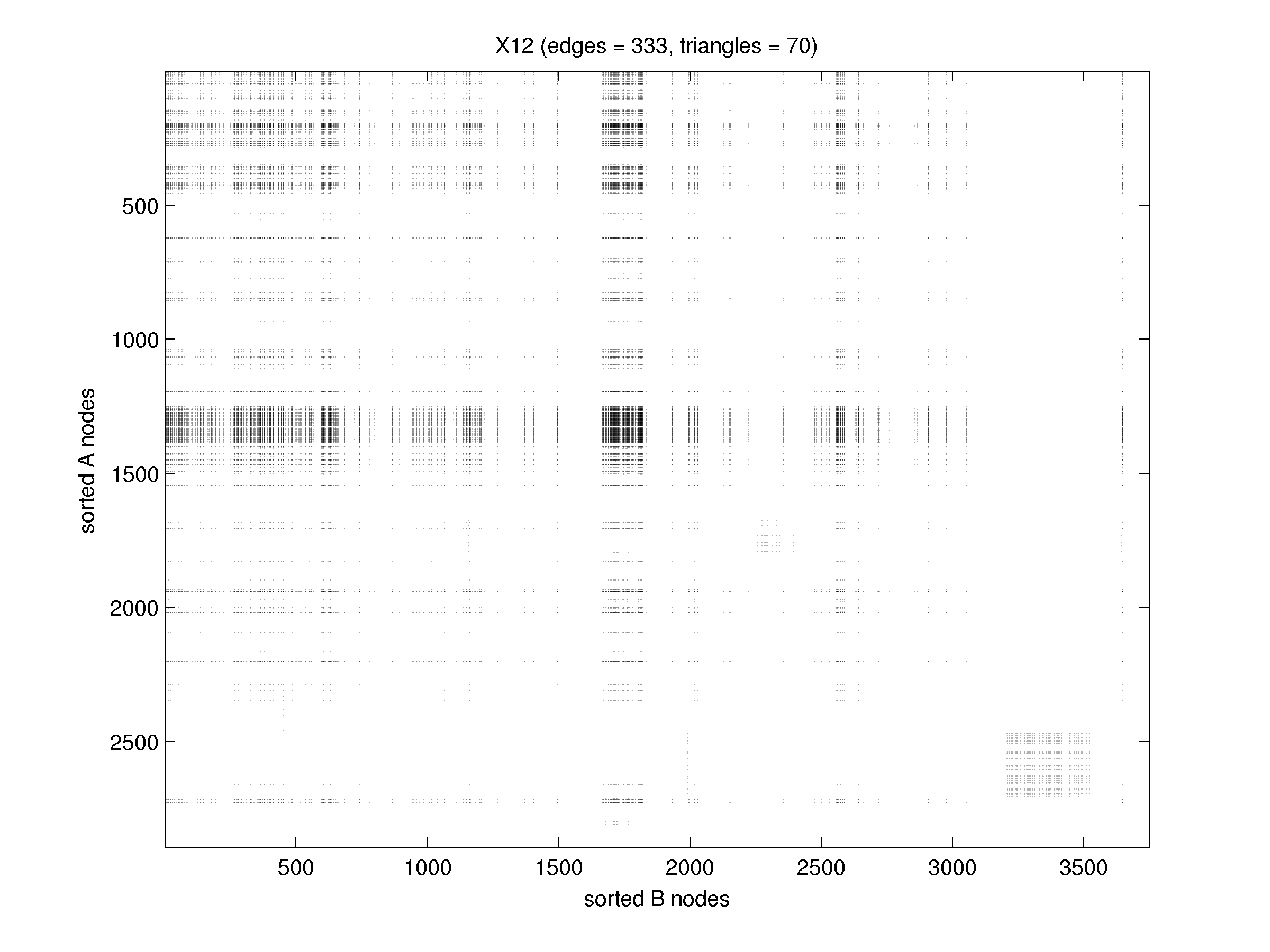}%
\label{fig:NAPA_X12}}
\hfil
\subfigure[Iteration 13]{\includegraphics[width=.19\textwidth]{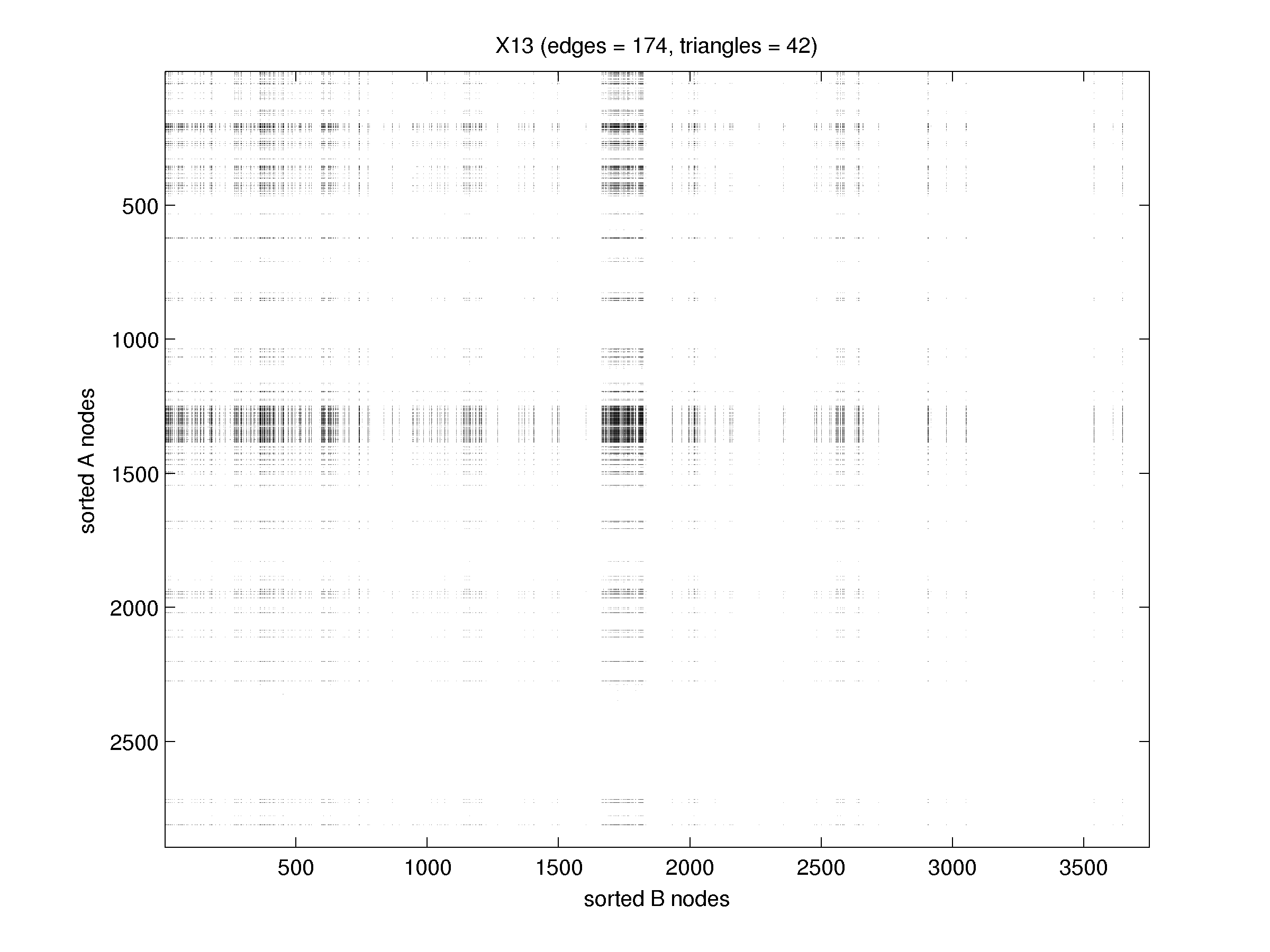}%
\label{fig:NAPA_X13}}
\hfil
\subfigure[Iteration 14]{\includegraphics[width=.19\textwidth]{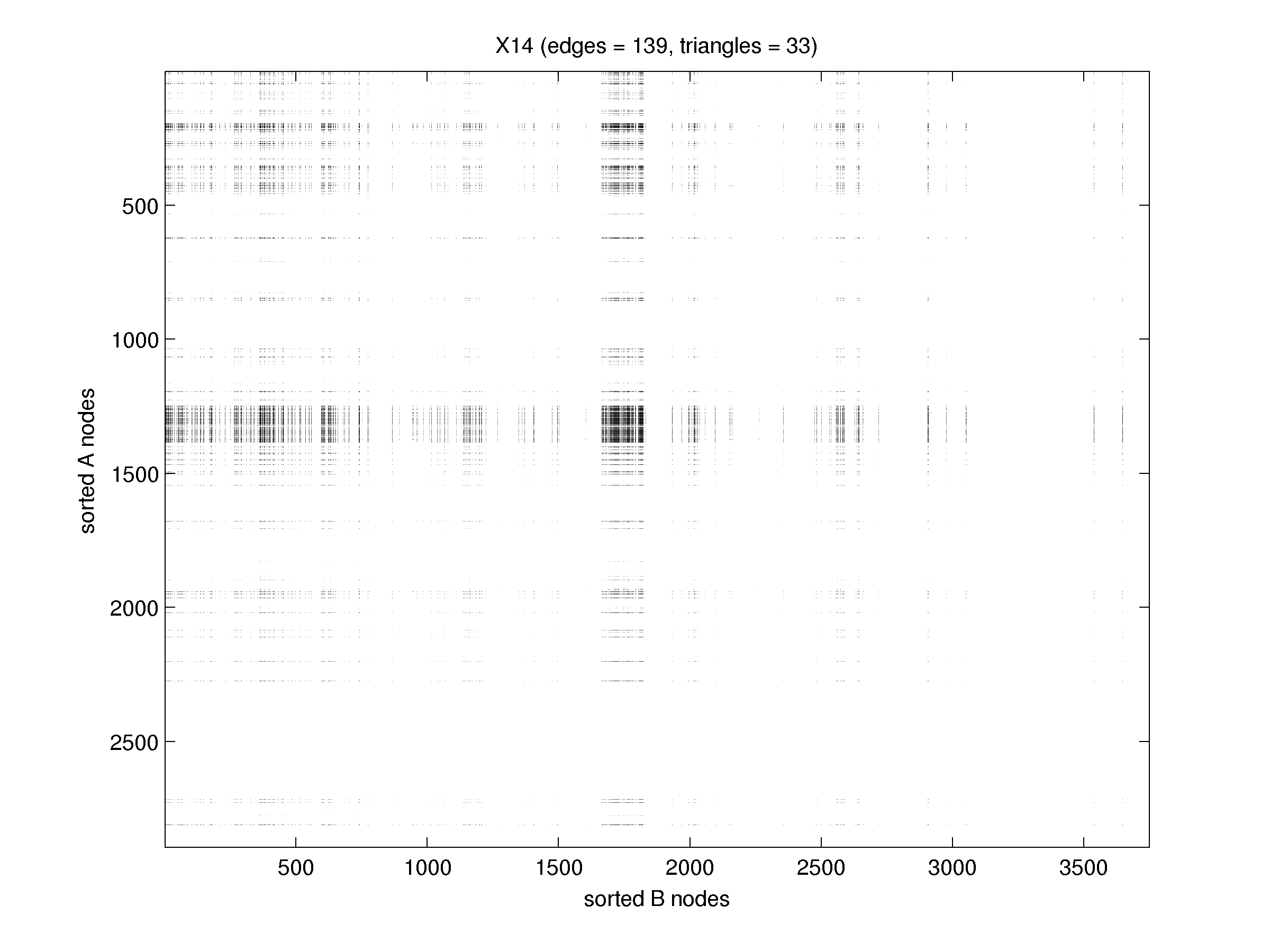}%
\label{fig:NAPA_X14}}
\hfil
\subfigure[Iteration 15]{\includegraphics[width=.19\textwidth]{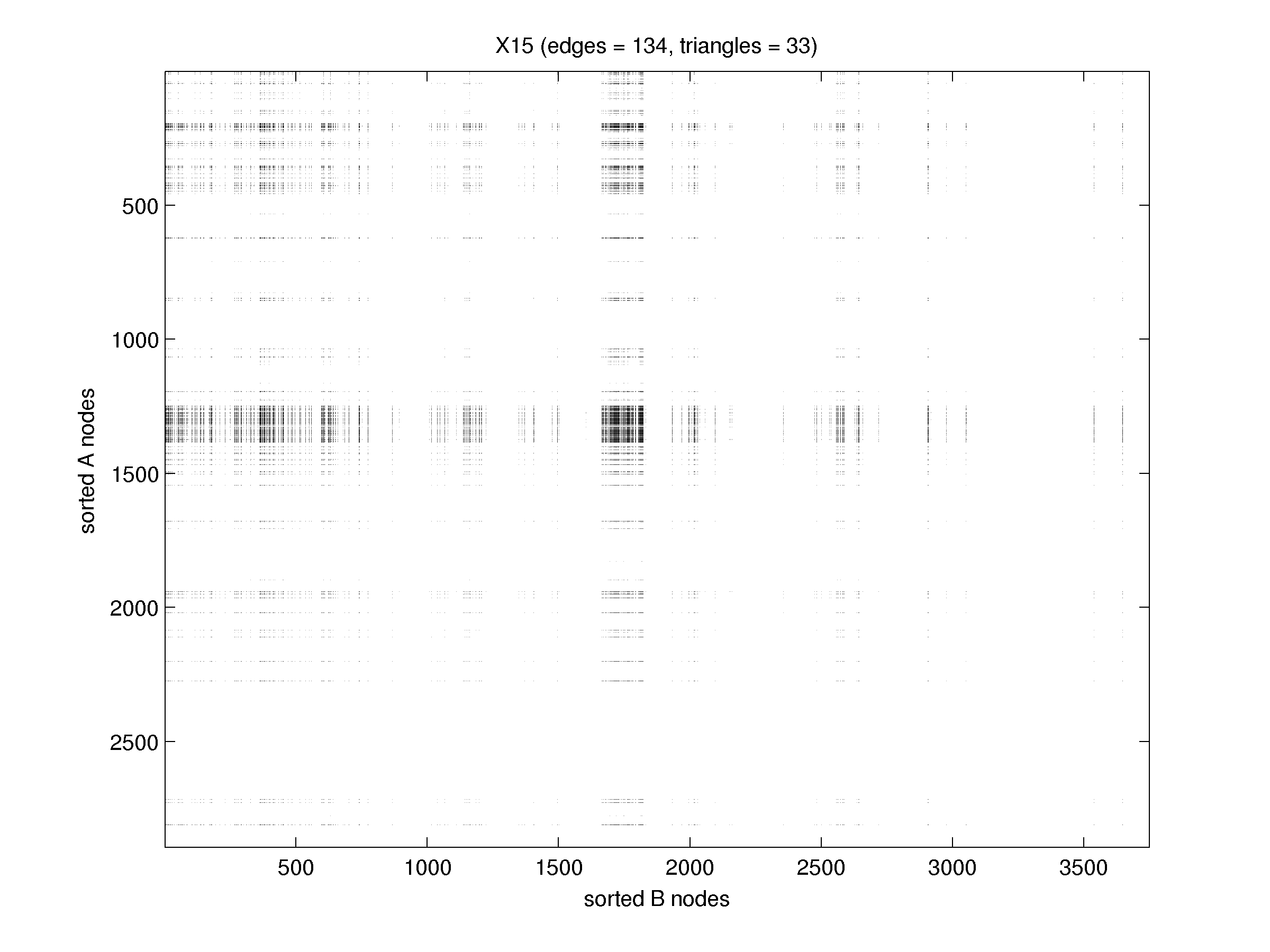}%
\label{fig:NAPA_X15}}
\hfil

\caption{The first 15 iterations of TAME applied to Family\_1 in the NAPAbench illustrated
as a matrix plot of iterates $\Vector{x}$ reshaped to a $\Matrix{X}$ where the true
orthologs lie on the diagonal. These illustrate how the best alignments result
from the information in the first few (2-5) iterations. }
\label{fig:NAPA_iterations}
\end{figure*}

A notable aspect of TAME's performance is that in almost all cases, the best solution occurred
within the first few iterations (2-3 in all cases we tried). The same characteristic is observed both for aligning the NAPAbench and 
real PPI networks. We note that since SS-HOPM does not have any means to internally avoid 
many-to-many mappings, the dominant eigenvector of $\Tensor{T}$ has a unique structure in which 
every node in one graph points to the most promising nodes in the other graph. In order to 
visualize this characteristic, we ran TAME over the Family 1 dataset in NAPAbench and visualized the 
structure of similarity matrix in each iteration. Figure~\ref{fig:NAPA_iterations} illustrates the 
first 15 iterations of the algorithm. We permuted rows and columns to highlight the orthologies as
the diagonal of the matrix. As such, the iterations start with all sequence similarities scattered around 
diagonal elements (Iteration 1), and many false positive off-diagonal pairs. As iterations continue, we start 
by finding a block diagonal structure (Iterations 2-5), representing triangle enriched regions in the networks.
As the process continues, one of the blocks emerges as the stationary point (Iterations 6-11).
Subsequent iterations localize around a solution induced by this block (Iterations 12-15). We are 
currently seeking theoretical characterizations of this behavior that may suggest improved methods.
For instance, it would be useful to avoid the transition to only one block that occurs during
Iterations 6-11.

\section{Concluding remarks and future work}

In this paper, we propose an alternative formulation of the network alignment problem that
uses higher-order substructures to drive the alignment process. We provide the 
necessary mathematical and algorithmic machinery for encoding different motifs using
tensors; and, as a proof of concept, use triangle motifs to show how the framework
can be applied to the network alignment problem. We show that our method
outperforms state of the art techniques in terms of the total number of triangles aligned (Figures ~\ref{fig:NAPA_NCV_tGS3} and ~\ref{fig:NCV_tGS3}), and identifies novel biological insights (Figure~\ref{fig:CoExp}).

Our method returns a set of topological scores that can be combined with many 
of the other ideas in the network alignment literature. For instance, the information
contained in the TAME iterates is largely orthogonal to the information produced by
methods such as GHOST. We believe it is likely that these different similarity
scores can be integrated -- perhaps by local features of the graph topology
to characterize their reliability -- potentially yielding a result that is
better than either. 

Our ongoing work is focused on optimizing the implicit kernel, enhancing mixing properties of 
sequence and topological similarities, extending the main iteration to simultaneous subspace 
iteration with nonnegative orthogonalization, combining motifs of different sizes into the 
optimization problem, and understanding the theoretical basis of the success
of the early iterations.

\ifCLASSOPTIONcompsoc
  \section*{Acknowledgments}
\else
  \section*{Acknowledgment}
\fi

This work is supported by the Center for Science of Information (CSoI), an NSF Science and 
Technology Center, under grant agreement CCF-0939370, as well as by NSF Grant BIO-1124962,
NSF Grant CCF-1149756, NSF Grant IIS-1422918, and the DARPA SIMPLEX Program.
This material is based upon work supported by the U.S. Department of
Energy, Office of Science, Office of Advanced Scientific Computing
Research, Applied Mathematics program.
Sandia National Laboratories is a multi-program laboratory managed and
operated by Sandia Corporation, a wholly owned subsidiary of Lockheed
Martin Corporation, for the U.S. Department of Energy's National
Nuclear Security Administration under contract DE--AC04--94AL85000.

\ifCLASSOPTIONcaptionsoff
  \newpage
\fi



%
\printbibliography

\begin{IEEEbiography}[{\includegraphics[width=1in,height
=1.25in,clip,keepaspectratio]{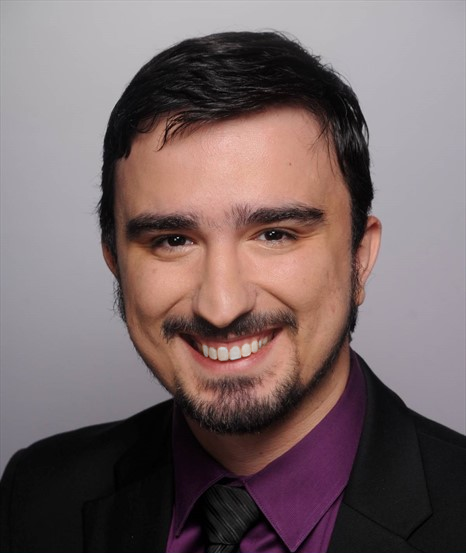}}]{Shahin Mohammadi}  received his Master's degree in Computer Science from Purdue University in December 2012 and is currently a Ph.D.~candidate at Purdue. His research interests includes computational biology, machine learning, and parallel computing. His current work spans different areas of Bioinformatics/ Systems Biology and aims to develop computational methods coupled with statistical models for computationally-intensive problems.
\end{IEEEbiography}

\begin{IEEEbiography}[{\includegraphics[width=1in,height=1.25in,clip,keepaspectratio]{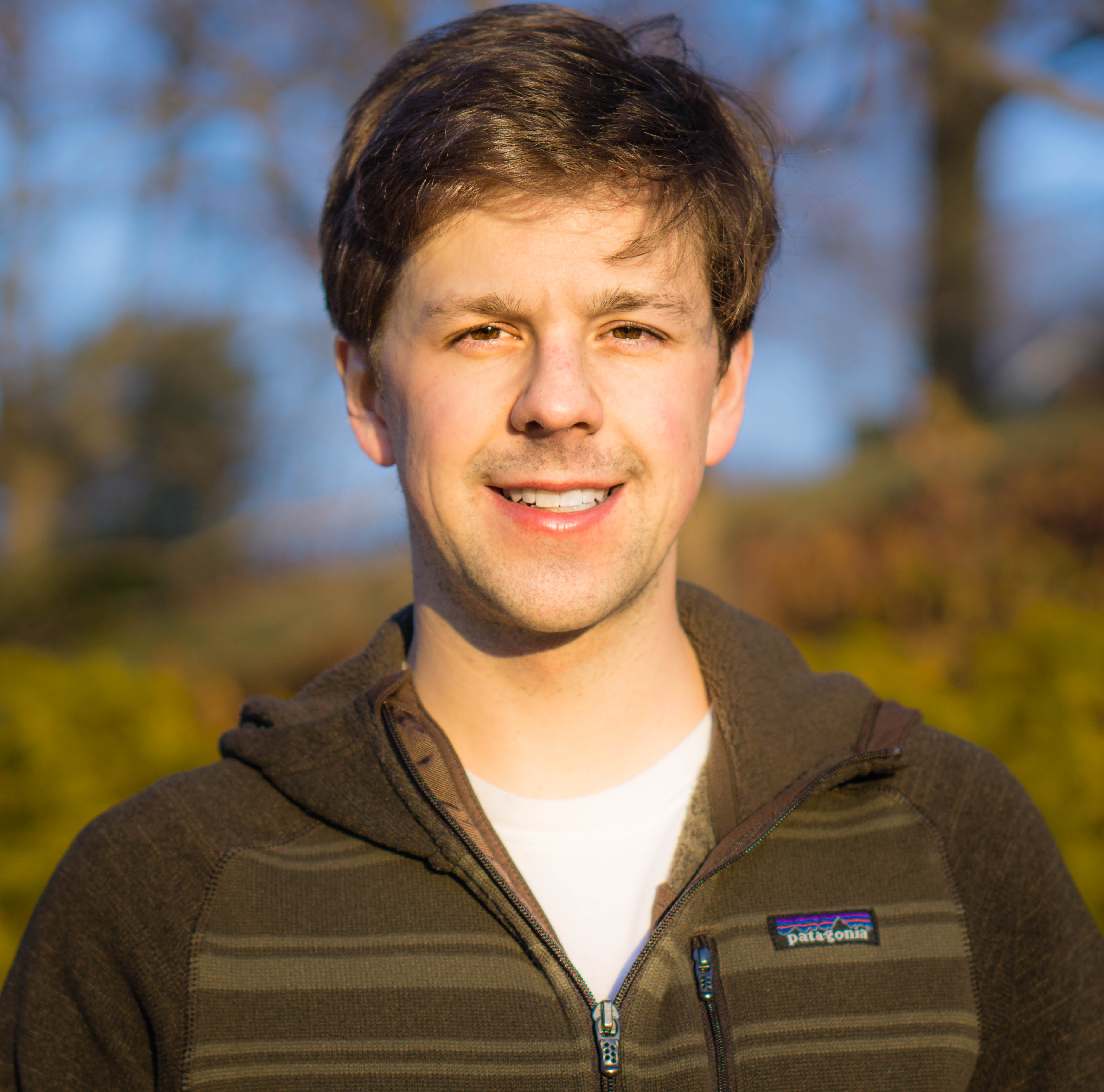}}]{David F.~Gleich} 
is an assistant professor of
Computer Science at Purdue University. His
research is on matrix computations, network
and graph algorithms, and parallel and 
distributed computing. He received a Bachelor
of Science degree from Harvey Mudd College, 
and a Ph.D.~from Stanford University.
He has been awarded a Microsoft Research
Graduate fellowship, the John von Neumann
postdoctoral fellowship, and an NSF 
CAREER award.
\end{IEEEbiography}

\begin{IEEEbiography}[{\includegraphics[width=1in,height=1.25in,clip,keepaspectratio]{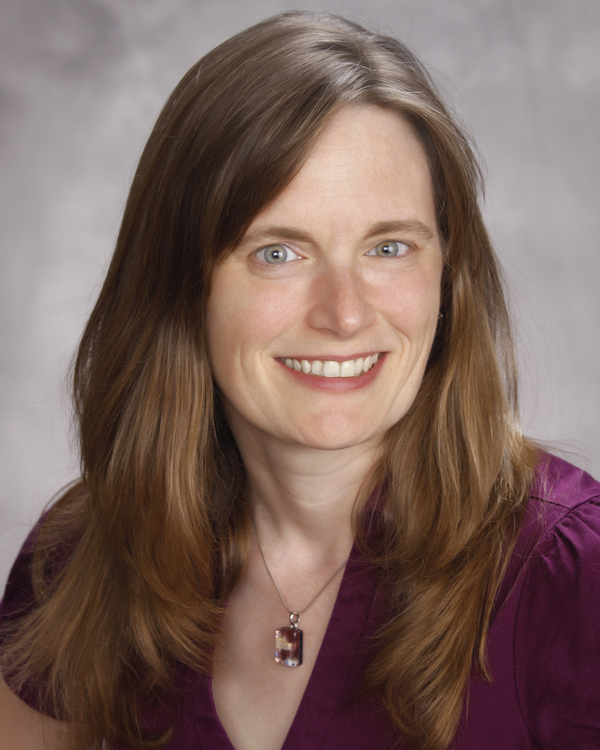}}]{Tamara G.~Kolda}
is a Distinguished Member of Technical Staff at Sandia National Laboratories in Livermore, California. Her research interests include multilinear algebra and tensor decompositions, graph models and algorithms, data mining, optimization, nonlinear solvers, parallel computing, and the design of scientific software. She received a Ph.D.~from the University of Maryland. 
She is a fellow of the Society for Industrial and Applied Mathematics (SIAM).
\end{IEEEbiography}

\begin{IEEEbiography}[{\includegraphics[width=1in,height
=1.25in,clip,keepaspectratio]{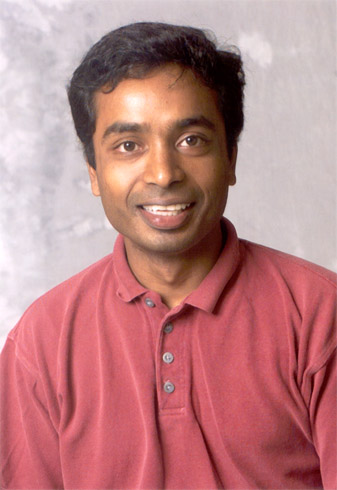}}]{Ananth Grama} received
a Ph.D.~degree from the University of Minnesota in 1996. He is
currently a Professor of Computer Sciences and Associate Director of
the Center for Science of Information at Purdue University.
His research interests span areas of
parallel and distributed computing architectures,
algorithms, and applications. On these topics, he
has authored several papers and texts. He is a member of the
American Association for Advancement of Sciences.
\end{IEEEbiography}

%
%
%
%

\end{document}